\documentstyle[preprint,tighten,aps,eqsecnum,epsfig]{revtex} 
\newcommand{\bnll}[1]{\begin{mathletters}\label{#1}\begin{eqnarray}}
\newcommand{\enll}{\end{eqnarray}\end{mathletters}}

\begin{document}

\draft


\preprint{\fbox{\sc version of \today}}

\def\lt{\raisebox{0.2ex}{$<$}}
\def\gt{\raisebox{0.2ex}{$>$}}

\title{Nuclear Liquid Drop Model with the Surface-Curvature Terms:
       New Perspectives for the Hyperdeformation Studies\thanks{This 
       work has been partly supported by the Polish Committee for Scientific 
       Research under Contract No. 2P03B~115~19 and by the Program  No. 99-95
       of Scientific Exchange between the IN$_2$P$_3$, France, and Polish 
       Nuclear Research Institutions.}
}
\author{
        K.~Pomorski$^{1,2}$ and J.~Dudek$^1$ \\
       {\it $^1$Institut de Recherches Subatomiques,
            IN$_2$P$_3$-CNRS/Universit\'e Louis Pasteur \\
            F-67037 Strasbourg Cedex 2, France}\\
       {\it $^2$Katedra Fizyki Teoretycznej, 
            Uniwersytet Marii Curie-Sk\l{}odowskiej,}\\
       {\it PL-20031 Lublin, Poland}
}


\maketitle

\begin{abstract}

Nuclear liquid drop model is revisited and an explicit introduction of the 
surface-curvature terms is presented. The  corresponding parameters of the
extended classical energy formula are adjusted to the contemporarily known
nuclear binding energies and fission  barrier heights. Using 2766 binding
energies of nuclei with $Z\geq 8$ and $N\geq 8$ it is  shown that the
performance of
the new approach is improved by a factor of about 6, compared to the previously
published liquid drop model results, in terms of both the masses 
(new r.m.s. deviation $<\delta M> = 0.698$ MeV) and the fission barriers 
(new r.m.s. deviation of the fission barriers of isotopes with $Z > 70$ is
$<\delta V_B> = 0.88$ MeV). 

The role of the curvature terms and their effects on the description of the
experimental quantities are discussed in detail; for comparison the parameters
of the more 'traditional' approaches are re-fitted taking into account the
nuclear masses known today and the performances of several variants of the 
model are compared.
The isospin dependence in the new description of the barriers is in a good 
agreement with the extended Thomas-Fermi approach; it also demonstrates a good 
qualitative agreement with the fission life-time systematics tested on the long 
chain of Fermium isotopes known experimentally.

The new approach offers also a very high stability in terms of the extrapolation
from the narrower range of nuclides to a more extended one - a property of
particular interest for the contemporary exotic beam projects: the corresponding
properties are illustrated and discussed.  

\end{abstract} 


%
%

%
%

\section{Introduction}
\label{Sec01}

It is more than sixty years by now since the first successful application of
the  charged liquid-drop model to describe the nuclear binding energies
\cite{We35,BB36}. Brilliant extensions of the Bethe-Weizs\"acker nuclear drop
concept by Meitner and Frisch \cite{MF39} and by Bohr and Wheeler  \cite{BW39}
have been obtained in 1939 and used to explain the nuclear fission  phenomenon.
Since then many papers have been devoted to the nuclear liquid drop  model
formalism and its improvements. Various new terms in the corresponding energy
expressions have been proposed but the basic concept of the charged  liquid
drop which could deform and fission remained valid. It is worth reminding at
this point that already in 1953 Hill and  Wheeler concluded on the basis of the
Fermi gas model, Ref.~\cite{HW53},  that a curvature dependent term
proportional to $A^{1/3}$ should exist in  the liquid drop energy functional.  
The curvature term was later studied in Ref.~\cite{GH69} where its magnitude 
was adjusted to the known at that time experimental fission barrier heights.

The macroscopic model description of the nuclear masses and, more generally,
the nuclear deformation energies using Strutisky-type approaches, plays
a very important role in the large scale nuclear energy calculations that
allow programming new important experiments such as e.g. on super- or on
still hypothetical hyper-deformations at high angular momenta. The existence
of the hyperdeformed nuclear configurations has been predicted on the basis
of the realistic large scale calculations - the same calculations that 
predicted the existence of several islands of the super-deformed nuclei. 
While the presence of the latter has been confirmed experimentally on several 
dozens of cases, there is so far no single convincing experimental 
evidence for the former. On the one hand, problems related to the increased 
{\em instrumental} sensitivity, when looking for manifestations of the 
hyperdeformed configurations had to be expected. On the other hand it is clearly
of importance to look for sources of possibly systematic and perhaps not so 
well controlled effects in the theoretical large scale calculations performed 
so far, and one may hope that by combining the two different types of efforts
some new steps forward will be possible.   

A possible mechanism to discuss that has not been taken into account in the 
large scale  calculations in question is related to the interpretation and more
generally to the mathematical representation of the classical nuclear surface
energy. Deformation-dependent classical energy expressions can be seen as 
functions of two groups of variables that describe, respectively, the nucleus 
itself, $(Z,N)$, and its shape represented by an ensemble of the deformation
parameters, here abbreviated to $\{ \alpha \}$. Typically, the surface energy
is written as a product $E_s(Z,N; \{ \alpha \} )$ =  $f(Z,N) g(\{ \alpha \})$,
where the first factor is usually parametrized by introducing a few adjustable
constants e.g. $f(N,Z)$ $=$  $p_0 + p_1 (N-Z)/(N+Z)$ or any other expression of
this type that is found performant; $p_1$ and $p_2$ are adjustable constants,
whose number does not need to be limited to 2. As it has been discussed already
by other authors, in a  more careful approach the nuclear surface energy can be
seen as contributed by  {\em two} different but related geometrical elements:
the numerical value of  surface area and the surface's average curvature (cf.
Eqs.~(\ref{Ecur}) - (\ref{Tayl}) below and the corresponding text). Such an
argument implies a different form of the surface energy expression: $E_s(Z,N;
\{ \alpha \} )$  $=$ $f_a(Z,N) g_a(\{ \alpha \})$ + $f_c(Z,N) g_c(\{ \alpha
\})$, where indices $a$ and $c$ refer to \underline{a}rea and 
\underline{c}urvature, respectively, and
where the deformation dependencies of $g_a$ and $g_c$ are different; moreover,
in the spirit of the classical nuclear energy models the corresponding factors
$f_a(Z,N)$ and $f_c(Z,N)$ are to be adjusted separately. By re-fitting all the
adjustable parameters of the classical energy expression to the experimental
masses of over two thousands of nuclei as well as on the fission barriers we
are going to look for the most performant parameterization to be used in
conjunction with the Strutinsky type formalism. In such an approach all the
terms including the surface energy term will be represented 'as optimally as
possible\footnote{ We expect that the {\em area} contribution $\sim f_a \cdot
g_a$ should be a dominating factor since the traditional liquid drop model
without explicit use of the curvature energy term performed  quite well
already; the surface-curvature term is expected to be smaller  and to play a
role of a correction. We will demonstrate that such a fit is  possible and
corresponds to a significant improvement of performance of the liquid drop
model formula.}  in  a global fit'.

Let us emphasize that interpretations of classical models that simulate
quantum properties of nuclei must not be associated too directly to the
notions of
non-existing objects such as {\em classical forces} supposed to act on the
nucleons e.g. in the vicinity of the nuclear surface. Strictly speaking,
classical  interpretations involving such forces that are related either to
the {\em surface tension} or {\em surface curvature} do not make much sense
for the strongly interacting Fermi systems. This implies in particular that in
the present context the corrective curvature term does not need to have any 
definite sign and the optimal effect can be obtained by taking, say, a larger
area-contribution accompanied by smaller curvature-contribution, or, to the
contrary, with a slightly smaller  area-contribution accompanied instead by a
larger curvature-term. In fact the surface energy contributions in the more
traditional approaches and in the present approach are close in terms of their
numerical  values but the present  approach that distinguishes between the 
{\em surface-area} contribution  and the {\em surface-curvature} contribution 
turns out to approximate the whole ensemble of the known nuclear masses and
barriers in a more flexible way. In particular the new r.m.s. deviation will be
shown to be $<\delta M> = 0.698 $ MeV compared to $<\delta M> = 0.732 $ MeV
within the traditional approach, and the new fission barrier r.m.s. deviation 
for nuclei with $Z > 70$, $<\delta V_B> = 0.88 $ MeV, compared to 
$<\delta V_B> = 5.58 $ MeV.

Several studies performed in the past, of the contributions coming from the 
surface curvature to the total energy, aimed at estimating its value using
a more elementary (microscopic) ideas about the nuclear interactions,
both for the finite nuclei and for the semi-infinite nuclear matter media.
Some of the corresponding papers are mentioned below; much more details
about that evolution can be found in the articles quoted therein. In particular,
using the energy density formalism of Ref.~\cite{Br69} combined with the
macroscopic formulation of the curvature energy expressions of Myers and
\'Swi\c{a}tecki, \cite{MS69}, Stocker, Ref.~\cite{St73}, pointed to the 
compatibility of the  curvature energy estimates coming from the two
approaches. Grammaticos, Ref.~\cite{Gr83}, using the Skyrme type functional,
but limiting himself to the terms of the order of $\hbar ^2$, 
was able to obtain what could be considered as a reasonable estimate
for the curvature energy, stressing however that the results are sensitive to
the details of the energy functional and pointing to the necessity of including
higher order terms. This has been done for instance in Ref.~\cite{SB88}, where 
also a comparison of the results of various calculations and estimates known
at  the time of publication can be found. However, the main results obtained by
the  authors, were compatible with the earlier theoretical predictions.  A more
distinct link between microscopic and macroscopic models was proposed in
Ref.~\cite{BG84}, where various terms of the droplet model were derived from 
the Skyrme interaction, in the framework of the extended Thomas-Fermi (ETF) 
model.
The problem of self-consistency, when approaching the issue of  the curvature
energy, has been addressed in \cite{DS93}; no major influence of this aspect of
the formalism on the final result has been found. Relativistic mean-field
theory within semi-classical approach has been applied, Ref.~\cite{CV93}, to
the semi-infinite nuclear matter concluding that the relativistic and the more
traditional methods give in essence compatible results. Extension to the
relativistic but quantum approaches has been studied in Ref.~\cite{CV96} with
the conclusion that also within the relativistic approaches the semi-classical
and fully quantum approaches give consistent, comparable results. Similar
physical goals but within relativistic Hartree approximation have been
approached in \cite{ES94} and sensitivity of the final result to the related
physical quantities such as the (in)compressibility coefficient and  nucleonic
effective-mass has been discussed. Also, a detailed, more recent  analysis of
the problem of the surface and curvature energies using Skyrme type
interactions but aiming principally at the astrophysical applications can be
found in Ref.~\cite{DH00}, see also references therein.

Let us stress that the above mentioned developments addressed first of all the
problem of an existence of relationships between the nuclear curvature energy
(terms in the total nuclear energy expression proportional to $A^{1/3}$) and a
microscopic representation of the nuclear forces, together with the role of
such elements and mechanisms as the order  of expansion in the extended
Thomas-Fermi model, type of the Skyrme forces,  comparison between the
semi-classical and the quantum calculation results, as well as the possible
influence of the relativistic effects. All these studies point coherently to
the result that the first order curvature coefficient should be of the order
of, typically, 5 to 15 MeV. At the same time most of the  more phenomenological
approaches, based directly on the {\em global fits} to the  experimental data
pointed to the value very close to zero. In fact, in several  studies the
corresponding term was often altogether neglected, and the  discrepancy
mentioned turned into a kind of a 'curvature anomaly' problem. This
contradiction (energy contribution that should exist according to most of the
physical/theoretical arguments {\em vs.} fits within the multi-parametric
macroscopic model that give almost vanishing contributions) can possibly be
indicative of a peculiarity contained in the macroscopic energy  dependence on
its parameters, that makes it impossible to extract the non-zero values of the
curvature terms from
the data on the masses the way these extractions were attempted: below, possible
alternative fitting procedures will be discussed and their results compared. 

\begin{figure}
  \begin{center}
  \leavevmode
  \epsfig{file=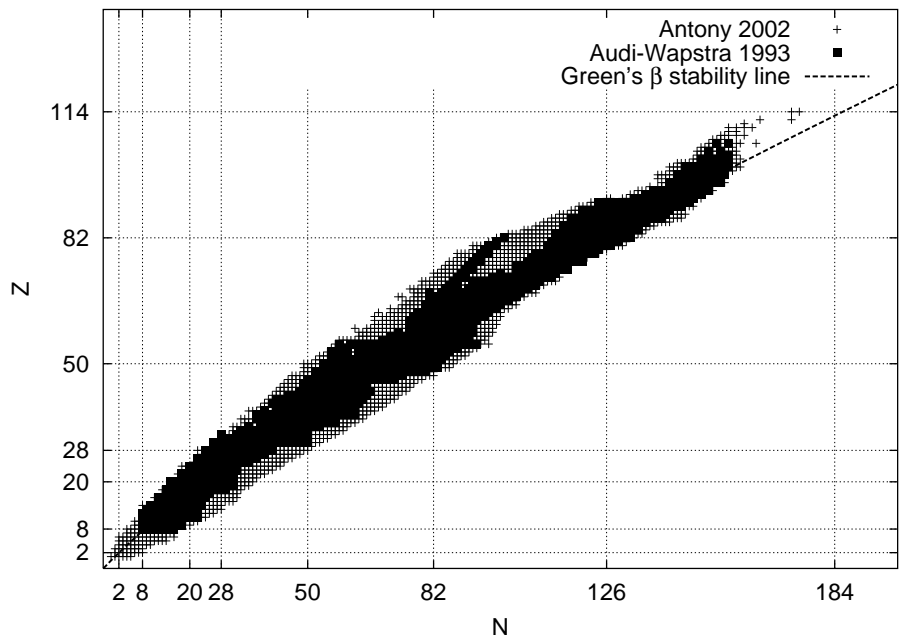, width=12.5cm, angle=00}
  \end{center}
  \caption{The chart of isotopes for which the experimental binding energies
          are known. The crosses correspond to data from the compilation of 
          Anthony \protect\cite{An01} while black squares to the data from 
          Ref. \protect\cite{AW93} on basis of which the analysis of Myers and 
          \'Swi\c{a}tecki \protect\cite{MS96} was done.}
  \label{fig01.lsd}
\end{figure}

There is also another group of studies that were focused more specifically  on
the calculations of the nuclear masses and/or the deformation dependence in the
classical energy expressions that, supplemented with the Strutinsky and pairing
quantum energy terms could be used for studying such problems as nuclear
fission, super- and hyper-deformation and more generally the shape coexistence
phenomena. A few years ago, a realistic Thomas-Fermi (TF) model has been 
developed by Myers and  \'Swi\c{a}tecki \cite{MS96}, that describes masses of
known nuclei with high  accuracy (the main ingredients of this model are
shortly recalled in  Appendix A). The corresponding r.m.s. deviation between the
experimental \cite{AW93} and theoretical binding energies for 1654 isotopes
amounts to 0.655 MeV only. In the last decade more than one thousand masses of
new isotopes have been  measured and in the new edition of the Strasbourg Chart
of Nuclides \cite{An01}  one can find 2766 binding energies of the isotopes
with the proton  and neutron numbers larger that Z=N=8 (cf.
Fig.~\ref{fig01.lsd}). The r.m.s. deviation of the TF estimates for these 2766
masses is 0.758 MeV and shows a high numerical precision of the model as well
as a good accuracy of the shell and  pairing energies obtained in
Ref.~\cite{MN95} that the TF model adopts. Fission barrier heights evaluated on
the basis of the Thomas-Fermi model \cite{MS96,MS99} are also in a rather good
agreement with the experimental data. 

A significant progress in the self-consistent methods has taken place in 
the recent years as well. For instance, the Hartree-Fock mass formula of 
Tondeur  {\em et al.}, Ref.~\cite{TG00}, that employs the effective $MSk7$ 
Skyrme interaction 
was able to reproduce the 1888 experimental binding energies with the r.m.s. 
deviation of 0.738 MeV. This r.m.s. deviation increases to only 0.828 MeV when one
makes the comparison with 2766 experimental masses taken from table \cite{An01}.

At present the self-consistent and the macroscopic-microscopic methods play
both their important roles in the nuclear structure calculations. While the
latter are very well suited for e.g. the 'automatic' large scale calculations
of the total nuclear energy surfaces, fission barriers, high spin properties,
shape-isomerism studies and/or numerous excited particle-hole configurations, 
the former are extremely useful in the detailed theoretical description of  the
nuclear states whose {\em global features} are already known. The simplicity
of  the macroscopic nuclear drop formalism together with the clear physical
meaning of its parameters add definitely to its attractiveness; it is easy to 
apply and thus frequently used in particular in estimating the fusion and 
fission cross sections in heavy ions reactions.

A particular  motivation for the present work is to obtain a new set of 
parameters of the liquid drop model adjusted to the up-to-date experimental 
masses  {\em and}  fission barriers, while taking a particular care of the 
surface-curvature
aspects of the model. This is of special importance when studying the exotic
nuclear shapes such as the nuclear hyper-deformation and/or the nuclear path to
fission (e.g. the bi-modal or more complex fission phenomena). The  nuclear
surface-curvature aspects were so far not analyzed very much in detail, at
least recently. As precise as possible a liquid drop model description
combined with the powerful shell energy description within the mean-field 
theories offers invaluable services.

A starting point of our analysis is the well known, 'traditional', liquid drop 
nuclear mass expression of Myers and \'Swi\c{a}tecki (MS-LD) \cite{MS67}.  This
expression was quite successful in reproducing the nuclear masses but it  is
known that in the light nuclei it overestimates the fission barrier heights  by
up to about 10 MeV \cite{KN79}; the MS-LD barriers are also higher than  those
evaluated by Sierk \cite{SI86} within the Yukawa-folded-interaction 
macroscopic model. 

It is of an obvious importance to assure the stability of the final result 
with respect to the cut off in terms of the number of multipoles used. All 
the fission-barrier heights presented in this paper were obtained by minimizing 
with respect to the deformation parameters $\beta_{\lambda}$ of even $\lambda$  
up to $\lambda_{max}=14.$  
In order to test the stability of our minimization procedure with respect to 
this cut off, we have performed additional test-minimizations using the 
Trentalange-Koonin-Sierk (TKS) family of shapes defined in Ref.~\cite{TK80}. 
The multipole and the TKS parameterizations clearly differ, yet the resulting
fission barriers almost coincide when number of the $\beta_{\lambda}$ 
parameters is sufficiently large. On the one hand, to obtain the same accuracy 
one needs typically twice as many multipoles as TKS deformation parameters. 
On the other hand we found out that the $\beta_{\lambda}$ parameterization 
is more stable than the TKS one when performing the numerical minimization 
of the potential energy surfaces (PES). Going beyond $\lambda_{max}=14$
does not change the final fission barrier results in the studied cases by
more than a couple of hundreds of keV for the highest barriers calculated
here, i.e. in the $A \sim 80$ mass range, while for the heavier nuclei the 
modifications are of the order of dozens of keV, an accuracy totally sufficient 
in the present context.

Direct calculations show that the Yukawa-folded-interaction 
model, which gives rather reasonable estimates of the fission barriers, is too
soft in directions perpendicular to the fission path especially at the large
nuclear elongation. It will be of great interest trying to combine (and we will
demonstrate that it is possible) an improved description of the fission
barriers together with a better description of the above mentioned stiffness
behavior within one single approach.

The paper is organized as follows: The actual version of the liquid drop model
used is described in Section \ref{Sect02}. In Section \ref{Sect03} we specify 
the way in which the parameters were determined and we present the
best sets of parameters for various variants of the LD models.  

Our results concerning the fission barriers are presented in 
Section \ref{Sect05}. The paper is summarized in Section \ref{Sect06}, 
where also an outlook of the planned applications of the obtained results
is presented.


\section{Liquid drop model and microscopic energy functional}
\label{Sect02} 

We are going to recapitulate briefly the main ideas of the leptodermous 
expansion \cite{MS74} of the energy-density functional in order to introduce 
the presentation of the role of the nuclear surface curvature-terms.


\subsection{General Considerations}
\label{Sect02a}            

The integral of the energy functional, the latter obtained e.g. by a 
self-consistent calculation, can be expanded into a power series of  
$A^{1/3}$ if one assumes that the energy-density function and the density of 
nucleons are diffused in the region of the nuclear surface (cf. e.g.
Ref.~\cite{BG84}).
Let us denote the nuclear radius by $R$, the nuclear surface thickness 
by $a$, and assume that $a/R \ll 1$ as well as that mass number $A$ is so 
large that $A^{-1/3} \ll 1$.
 
Let $\hat H$ denote the many-body Hamiltonian and $\Psi$ the corresponding 
many-body ground state wave function of a nucleus. The one-body density of 
the nuclear matter in the nucleus can be expressed as
\begin{equation}
      \rho  
      = 
      A \int \int \ldots \int 
      \Psi^{\star}\Psi \, 
      d\tau_2 \cdots d\tau_{A} \,,
                                                                     \label{E1}
\end{equation}
and the corresponding energy density as
\begin{equation}
      \eta  
      = 
       \int \int \ldots  \int  
      \Psi ^{\star} \hat{H} \Psi \,
      d\tau_2\cdots d\tau_{A} \,.
                                                                     \label{E2}
\end{equation}
The total energy of the nucleus is equal to the volume integral of the 
energy density  
\begin{equation}
       E  
       = 
       \int_{\rm V} \eta\, d^{\,3}{\bf r}\, ; 
                                                                     \label{E2a}
\end{equation}
it can be decomposed into the sum of the volume term and of another term 
that contains an integral over the nuclear surface $\Sigma$: 
\bnll{E3}
       E  
       = 
       \int_{\rm V} (\eta + b_{\rm vol}\,\rho
                          - b_{\rm vol}\,\rho ) \, d^{\,3} {\bf r} 
     & = & b_{\rm vol} \,\int_v \rho            \, d^{\,3} {\bf r} 
       + 
       \int^{}_{\rm V} (\eta  - b_{\rm vol} \,\rho )\, d^{\,3} {\bf r} 
                                                                  \label{E3a} \\
     & = & b_{\rm vol} A 
       + 
       \int^{}_{\Sigma } d\sigma  \int^{\infty }_{0}
                          (\eta  - b_{\rm vol}\,\rho )\, dr_\perp \,. 
                                                                  \label{E3b}
\enll
The difference under the last integral in Eq.~(\ref{E3b}) can be expected to 
partially cancel out in the nuclear interior, the most significant contribution 
coming from the surface region. The remaining expression can be interpreted as 
the integral over the nuclear surface $\Sigma$ of the surface-tension $\gamma$, 
the latter formally defined by
\begin{equation}
      \gamma 
      \equiv 
      \int^\infty_0 (\eta - b_{\rm vol}\,\rho ) \, dr_\perp \,.
                                                                     \label{E5}
\end{equation}                       
Let us introduce the principal radii $R_1$ and $R_2$ that are associated
locally to any point on the nuclear surface $\Sigma$. In geometry, the local
properties of surfaces are conveniently expressed in terms of the
first order curvature $\kappa$ and the second order (Gauss) curvature $\Gamma$
defined through
\begin{equation}
      \kappa  
      = 
      \frac{1}{R_1} + \frac{1}{R_2}
      \quad {\rm and} \quad
      \Gamma  
      = 
      \frac{1}{R_1 \cdot R_2} \,,
                                                                     \label{E6}
\end{equation}                       
respectively. These {\em local} quantities can be shown to underly important
and interesting in the present context {\em global} properties. Indeed, it
is easy to see that in geometrical terms, the volume $\mathcal{V}$ enclosed 
by a given surface $\mathcal{S}$, the surface itself and the implied average 
curvature $\mathcal{L}$ are directly correlated through the equation of the 
surface since one may write
\bnll{Ecur}
      \mathcal{V}
      &=&
      \frac{1}{3} \int_{\mathcal{S}} d\vec \sigma \cdot \vec{r}\,, 
                                                              \label{Ecura} \\
      \mathcal{S}
      &=&
      \int_{\mathcal{S}} d\sigma\,, 
                                                              \label{Ecurb} \\
      \mathcal{L}
      &=& 
      \int_{\mathcal{S}} d\sigma\, \bigg( \frac{1}{R_1} + \frac{1}{R_2} \bigg) 
       \,. 
                                                              \label{Ecurc}
\enll
An instructive next step can be obtained by considering a family of
some special surfaces, the one-parameter Steiner sheets $\{ {\mathcal{S}}(s) \}$,
having the property of being universally equidistant. More precisely, for any
two values of parameter $s$, say, $s_1$ and $s_2$, the corresponding
normal distances are constant and equal $\vert s_2 - s_1 \vert$. 
For such surfaces it can be shown that
\begin{equation}
       {\mathcal{L}} (s)
       =
       \frac{d {\mathcal{S}} }{d s}
       =
       \frac{d^2 {\mathcal{V}} }{d s^2}
                                                              \label{Ecur1}
\end{equation}
with the consequence, as remarked in Ref.~{\cite{GH69}}, that for the Taylor
expansions, e.g. at $s=0$, one finds 
\bnll{Tayl}
       {\mathcal{V}} (s)
      &=&
       {\mathcal{V}}_0
       +
       {\mathcal{S}}_0 \, s
       +
       \frac{1}{2} {\mathcal{L}}_0 \, s^2 + \ldots
                                                              \label{Tayla} \\
       {\mathcal{S}} (s)
      &=&
       {\mathcal{S}}_0
       +
       {\mathcal{L}}_0 \, s + \ldots \, ,
                                                              \label{Taylb}
\enll
where ${\mathcal{V}}_0$, ${\mathcal{S}}_0$ and ${\mathcal{L}}_0$ are constants,
equal to the values of the functions in Eq.~(\ref{Ecur1}) at $s=0$.
Relations of an analogous mathematical structure are used more generally 
in the LD approximations and also below in this article. In the form 
(\ref{Tayl}) they show that the average curvature indeed characterizes
the volume and the surface within Taylor expansion of the second and first
order, respectively, and that this quantity is indeed a natural geometric
feature to introduce in the related physical considerations\footnote{The
surfaces used in the macroscopic description of nuclei are in general
not strict examples of Steiner sheets. Yet, in many cases the regular surfaces
used are expected to behave in a very similar manner; the very fact that
the expansions used in physics resemble those in Eq.~(\ref{Tayl}) and work well 
in physical applications strongly suggests that this is in fact the case. }.
The above relations are suggestive of yet another important possibility
related to the question of the sign of the curvature term that has been for 
some time also a discussed issue.
The surface effect represented by the Taylor expansion of ${\mathcal{S}}(s)$
in Eq.~(\ref{Taylb}), is a sum of two terms, and as it stands suggests that
both ${\mathcal{S}}_0$ and ${\mathcal{L}}_0$ are uniquely defined constants.
In the applications, however, the procedure followed is different: the 
expressions that mathematically resemble the above relation(s) are still used
but the parameters are {\em fitted} using {\em Steiner-like} surfaces rather 
than using Steiner's theorem. Consequently, whenever ${\mathcal{S}}_0$ is
found slightly larger, the corresponding  ${\mathcal{L}}_0$ will provide a
compensation, including a possibility that ${\mathcal{L}}_0 < 0$. We believe
that in the present context of the Taylor-expansion type of analysis it would 
have been inappropriate to associate a definite sign of such
a contribution as 'more physical than the other' - in contrast - the sign of
the whole surface contribution could be attributed physical sense.

It can be shown that for the surface of sufficient regularity the {\em average} 
Gauss curvature defined in analogy to the average first order curvature through 
the corresponding surface integral satisfies
\begin{equation}
      \int_{{\mathcal{S}}} d\sigma \, \bigg( \frac{1}{R_1 \cdot R_2} \bigg) 
      = 
      4 \, \pi \, ,
                                                                     \label{Gau}
\end{equation}
independently of the actual shape; this feature will have interesting
consequences for the reproduction of the nuclear masses (see below). 
 
One can expect that the local surface tension depends on the diffusivity
of the surface region (represented by the diffusivity parameter $a$)
as well as on the two curvatures. For dimensionality reasons it will be
convenient to parameterize this dependence as
\begin{equation}
      \gamma 
      = 
      \gamma \,(a\kappa, a^2 \Gamma ) \,.
                                                                     \label{E6a}
\end{equation} 
The above function can be decomposed into a Taylor series around the 
argument values $a \kappa = 0$ and $a^2 \Gamma=0$ and we obtain
\begin{equation}
      \gamma (a\kappa, a^{\,2} \Gamma) 
      =  
      \gamma^{(0)}
      + 
      \gamma^{\,\prime}_{\kappa} \; \kappa \; a 
      +
      \frac{1}{2} \gamma^{\,\prime \prime}_{\kappa\kappa}\,\kappa^{\,2}\;a^{\,2} 
      + 
      \gamma^{\,\prime}_{\Gamma} \, \Gamma \, a^{\,2} + \ldots \;,
                                                                     \label{E7}
\end{equation}
where $\gamma^{(0)}$, $\gamma^{\,\prime}_{\kappa}$, 
$\gamma^{\,\prime \prime}_{\kappa\kappa}$ and $\gamma^{\,\prime}_{\Gamma}$  
are constants.
Inserting the latter expression into the surface integral in Eq.~(\ref{E3b})
we may transform the surface contribution to the energy of the system 
to the following form:
\begin{eqnarray}
      \int^{}_{\rm V}
      (\eta  - b_{\rm vol} \,\rho) \, d^{\,3} {\bf r}  
      = 
      \gamma^{(0)} \int_{\Sigma } \, d\sigma 
      + 
      \gamma^{\,\prime}_{\kappa } \; a
      \int^{}_{\Sigma } \kappa \, d\sigma  
      +
      \frac{1}{2} \gamma^{\,\prime \prime}_{\kappa \kappa } \; a^{\,2}
      \int_{\Sigma } \kappa^2 d\sigma 
      + 
      \gamma^{\,\prime}_{\Gamma} \; a^{\,2} \int_{\Sigma } \Gamma \, d\sigma  
      + 
      \ldots \;.                                                   \nonumber \\
                                                                     \label{E8}
\end{eqnarray}
Since for the nuclear surfaces we have $\int_{\Sigma } \, d\sigma \sim A^{2/3}$ 
while $\kappa \sim A^{-1/3}$ and $\Gamma \sim A^{-2/3}$ it follows that
the terms appearing in the last relation can be alternatively parametrized as
\bnll{E9}
      &&\gamma^{(0)} \int_{\Sigma } \, d\sigma 
        \quad \to \quad  
         b_{\rm surf} \, A^{2/3} \, B_{\rm surf}\,, 
                                                                 \label{E9a} \\
      && \gamma^{\,\prime}_{\kappa } \; a 
         \int^{}_{\Sigma } \kappa  \, d\sigma 
         \quad \to \quad 
         b_{\rm cur} \, A^{1/3} \, B_{\rm cur}\,, 
                                                                 \label{E9b} \\
      && \frac{1}{2} \gamma^{\,\prime \prime}_{\kappa \kappa } \; a^{\,2}
         \int_{\Sigma } \kappa^2 \, d\sigma 
         + 
         \gamma'_{\Gamma} \; a^{\,2}
         \int_{\Sigma } \Gamma \, d\sigma 
         \quad \to \quad  
          b_{\rm curG} \, A^0 \,, 
                                                                 \label{E9c}
\enll
where we have inserted an explicit dependence of various terms on powers of
$A^{1/3}$ as well as the corresponding proportionality coefficients. The nuclear
deformation-dependent functions  $B_{\rm surf}$ and $B_{\rm cur}$ in relations
(\ref{E9a}) and (\ref{E9b}) are defined, respectively, as ratios of the 
surface and the mean-curvature, calculated at a given deformation, to the 
corresponding values at the spherical shapes. The Gaussian curvature energy 
(the last term in relation (\ref{E9c})) is deformation independent but may 
introduce 
an important dependence in terms of isospin when fitting the related liquid drop
parameters to the experimental masses; the first term in the correspondence
relation (\ref{E9c}) is small and its dependence on deformation, which we shall 
neglect later, can be found in Ref.~\cite{BG84}, cf. Eq.~(5.21) in the above 
reference.

The nuclear part of the total energy of a 
nucleus  can thus be given by the following final expression 
\begin{equation}
      E 
      = 
      b_{\rm vol}A 
      + 
      b_{\rm surf}A^{2/3} 
      +
      b_{\rm cur}A^{1/3} 
      + 
      b_{\rm curG}A^{0} + \ldots \;;
                                                                 \label{E10}
\end{equation}
the Coulomb part will be introduced later.


\subsection{Particular Case: Spherical Nuclei}
\label{Sect02b}            

\begin{figure}
  \begin{center}
    \leavevmode
    \epsfig{file=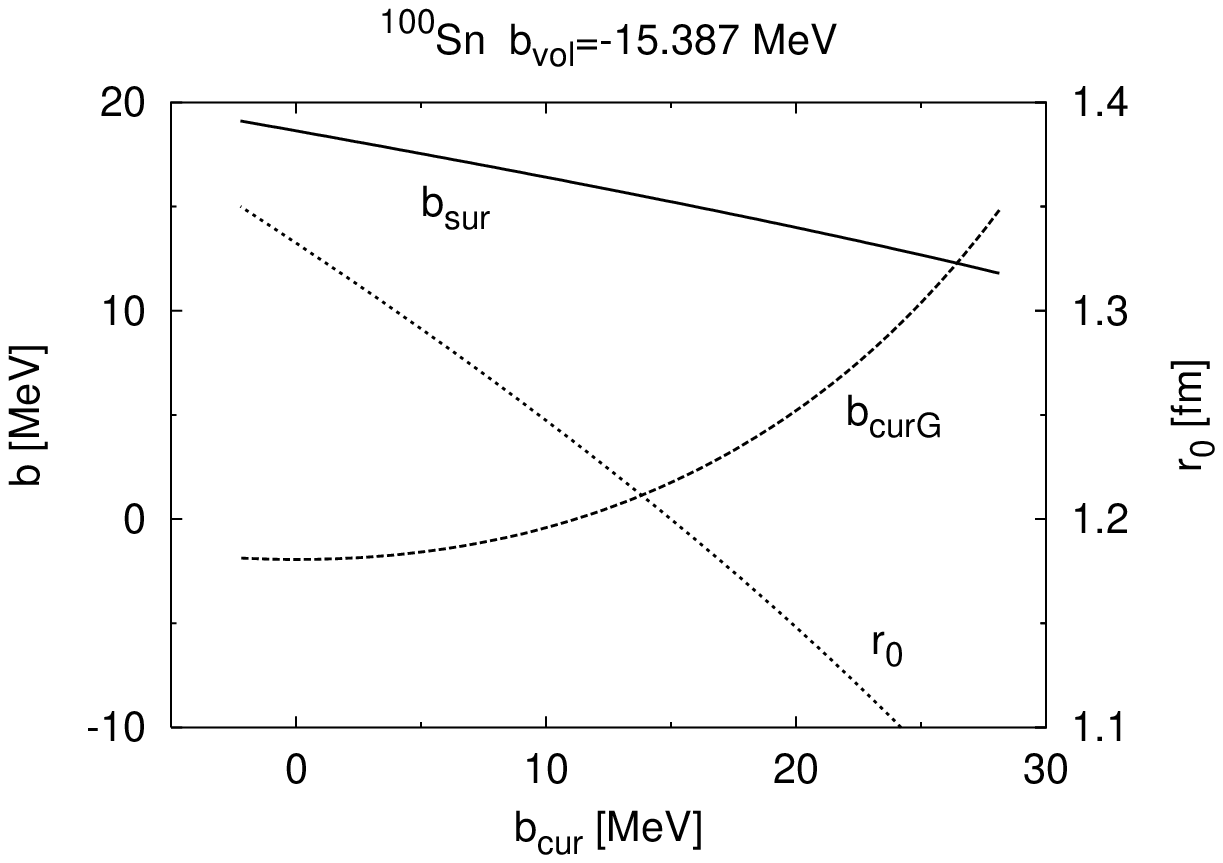, width=12.5cm, angle=00}
    \epsfig{file=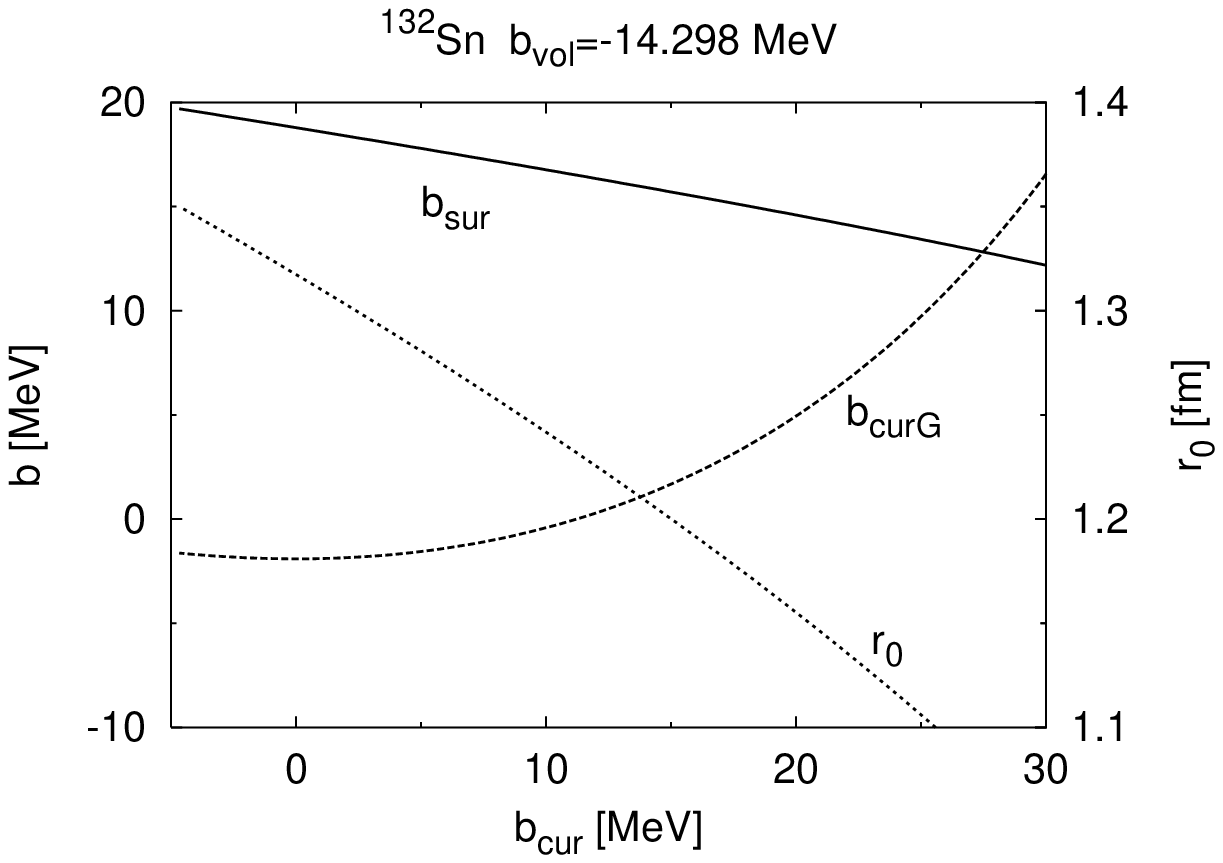, width=12.5cm, angle=00}
  \end{center}
  \caption{Interplay between the total value of the first order curvature 
            ($b_{\rm cur}$), the surface ($b_{\rm surf}$) and the second order 
            (Gauss) curvature ($b_{\rm curG}$) terms evaluated in the 
            leptodermous expansion around $R= r_0 A^{1/3}$ of the ETF energy 
            functional obtained with the Skyrme forces (SkM$^*$) for $^{100}$Sn 
            (top) and $^{132}$Sn (bottom). The corresponding values of $r_0$ 
            refer to the right-hand side ordinate axis. }
  \label{fig02.lsd}
\end{figure}

It is instructive to study the properties of expression (\ref{E10}) in the case 
of spherical nuclei. In this case the second integral in (\ref{E3b}) can be 
rewritten as follows:
\bnll{E11}
      E 
     &=& 
     b_{\rm vol} A
      + 
      \int_V (\eta - b_{\rm vol}\rho) \, d^{\,3} {\bf r} 
                                                                \label{E11a} \\ 
     &=& 
     b_{\rm vol} A 
      + 
      \int_\Sigma R^{\,2} \, d\Omega 
      \int_0^{\infty} (\eta -b_{\rm vol}\rho) \frac{r^{\,2}}{R^2} \, dr \;,
                                                                \label{E11b}
\enll
where $R$ (usually represented as $R=r_0 A^{1/3}$) is the radius of the 
spherical surface. Making use of the identity:
\begin{equation}
       {r^{\,2} \over R^{\,2}} 
       = 
       1 + \frac{2}{R}(r-R) + \frac{1}{R^{\,2}} (r-R)^{\,2} \;,
                                                                \label{E12}
\end{equation}
one can rewrite the remaining surface-related integral and transform the energy
expression as follows 
\bnll{E13}
      E            
     &=&  
      b_{\rm vol} A                                             \label{E13a} \\
     &+& 
     \int_\Sigma R^{\,2} d\Omega 
     \int_0^{\infty} (\eta -b_{\rm vol}\rho) \, dr
                                                                \label{E13b} \\
     &+& 
     \int_\Sigma 2R \, d\Omega 
     \int_0^{\infty} (\eta -b_{\rm vol}\rho) \, (r-R) \, dr     \label{E13c} \\
     &+& 
     \int_\Sigma d\Omega 
     \int_0^{\infty} (\eta -b_{\rm vol}\rho) \, (r-R)^{\,2} \, dr \; .
                                                                \label{E13d}
\enll
Above, expressions (\ref{E13b}), (\ref{E13c}) and (\ref{E13d}), contain terms
proportional to $R^2$, $R^1$ and $R^0$, respectively, thus at the same time, 
proportional to $A^{2/3}$, $A^{1/3}$ and $A^0$. In the present context they
should be interpreted as representing the surface, curvature and Gauss-curvature
contributions, correspondingly.
The nuclear part of the total energy of a spherical nucleus can thus be written 
down as
\begin{equation}
      E 
      =
      b_{\rm vol} A
      + 
      \underbrace{4\pi R^2 \cdot {\mathcal{I}}_0}_{\displaystyle b_{\rm surf}A^{2/3}} 
      + 
      \underbrace{8\pi R \cdot ({\mathcal{I}}_1-{\mathcal{I}}_0 R)}
                _{\displaystyle b_{\rm cur}A^{1/3}}
      +
      \underbrace{4\pi \cdot ({\mathcal{I}}_2 - 2 R {\mathcal{I}}_1 
                                              + R^2 {\mathcal{I}}_0)}
                _{\displaystyle b_{\rm curG}A^{0}}, 
                                                                \label{E14} 
\end{equation}
where the above mentioned correspondence relations are marked explicitly, and 
where
\bnll{E15}
     {\mathcal{I}}_0 
    &=& 
     \int_0^{\infty}(\eta - b_{\rm vol}\rho) \, dr \;,          \label{E15a} \\
     {\mathcal{I}}_1 
    &=& 
     \int_0^{\infty}(\eta - b_{\rm vol}\rho)\, r \, dr \;,      \label{E15b} \\
     {\mathcal{I}}_2 
    &=& 
     \int_0^{\infty}(\eta - b_{\rm vol}\rho)\,r^{\,2} dr \;, 
                                                                \label{E15c}
\enll
are radial moments associated with the nuclear surface layer.
Relation (\ref{E14}) allows to find, among others, a dependence between the 
curvature, surface and Gauss-curvature terms that follow from the ETF method. 
To start, $\eta$ and $\rho$ are calculated from  
(\ref{E1})-(\ref{E2}) using ETF method with Skyrme (SkM$^*$) forces of 
\cite{BQ82} wherefrom the integrals ${\mathcal{I}}_0$, ${\mathcal{I}}_1$ and 
${\mathcal{I}}_2$ are obtained. 
Next we proceed as follows: from Eq.~(\ref{E14}), for each predefined value of 
$b_{\rm cur}$ we write down equality  
$b_{\rm cur}A^{2/3}$ = $8\pi R \cdot ({\mathcal{I}}_1-{\mathcal{I}}_0 R)$
and, given ${\mathcal{I}}_1$ and ${\mathcal{I}}_0$, we deduce the implied 
$R$-value. The latter quantity known, we insert it into 
$b_{\rm surf}A^{2/3}$ = $4\pi R^2 \cdot {\mathcal{I}}_0$ and
$b_{\rm curG}A^{0}$ = $4\pi \cdot 
({\mathcal{I}}_2 - 2 R {\mathcal{I}}_1 + R^2 {\mathcal{I}}_0)$ and deduce
$b_{\rm surf}$ and $b_{\rm curG}$. Results of these operations are presented
in Fig.~\ref{fig02.lsd} for $^{100}$Sn (top) and $^{132}$Sn (bottom) tin 
isotopes. It is seen from the figure that the surface energy becomes smaller 
when the curvature constant is growing. The radius constant corresponding to 
the leptodermous expansion and evaluated via relation 
$R=r_0 A^{1/3}$ is marked on the right-hand side y-axis.

Finally let us observe the following interesting property. If we choose radius 
parameter $R$ in such a way that the Gauss curvature term [cf. the last term 
in Eq.~(\ref{E14})] is minimal i.e.: 
\begin{equation} 
      R 
      =
      \frac{{\mathcal{I}}_1}{{\mathcal{I}}_0}\;,
                                                                \label{E16}
\end{equation}
then the first order curvature term [the second one in Eq.~(\ref{E14})] is 
equal to zero. Even though we are not going to impose this condition in
what follows, it is instructive and helpful in analyzing the related
description of the nuclear masses to know about the existence of the above 
correlation, especially when examining the role of the second-order (Gauss)
curvature term.

The above observations confirm and illustrate the fact that the curvature terms 
in the nuclear energy are strictly related to the surface term as suggested in
Sect.~\ref{Sect02a} within a general introduction and that one can not discuss 
them separately. An increase of the first order curvature energy causes a 
decrease of the surface tension and {\em vice versa}. These observations will
have consequences for the fitting procedures applied below.

                              
\section{Fitting the liquid drop model parameters}
\label{Sect03}

Our aim is to find the parameters of the liquid drop model which correspond
to the leptodermous expansion of the nuclear energy [see Eq.~(\ref{E10})] and 
the Coulomb energy of a charged nuclear drop with a diffused surface.
We are going to consider separately four variants of the liquid drop model:
a. The one of Myers and \'Swi\c{a}tecki, Ref.~\cite{MS67}, with its original
fit of parameters, referred to as MS-LD;
b. Similar to the above but with the newly fitted constants, the fit using
the contemporary experimental data set and the microscopic energy
corrections\footnote{To be able to compare our results with those of the quoted
authors, the microscopic energy corrections for the lightest nuclei, more
precisely, those with $Z < 29$ and $N < 29$, were taken from \cite{MS96};
those for all heavier nuclei from \cite{MN95}.} - 
this variant is referred to as LDM;
c. The modernized version of the liquid drop model that contains the 
Gauss-curvature term, is in the following referred to as 'new', NLD, and 
d. Similar to the above but containing the deformation-dependent
first-order curvature term - this variant referred to as {\em Lublin-Strasbourg}
version of the nuclear {\em Drop} energy formula, abbreviated to LSD.

We begin by presenting the main features of the liquid drop energy dependence
on the surface-curvature terms.


\subsection{Liquid Drop Masses with Curvature Terms: Characteristic Features}
\label{Sect03a}            

We assume, in accordance with the usual rules of the liquid drop model
approaches, that the mass of an atom with Z protons, Z electrons and 
N neutrons is described by the following relation (cf. Refs. \cite{MS67,MS96}):  
\begin{eqnarray}
      M(Z,N;{\rm def} ) 
     &=&  
      Z M_{\rm H} 
      + 
      N M_{\rm n} 
      - 
      0.00001433 \, Z^{2.39}                                      \nonumber \\ 
     &+& 
      b_{\rm vol}\;\,(1 - \kappa_{\rm vol} \; I^2\,)\,A           \nonumber \\
     &+&
      b_{\rm surf}\,(1 - \kappa_{\rm surf} I^2\,)\,A^{2/3} 
      B_{\rm surf}({\rm def}) 
                                                                  \nonumber \\
     &+&
      b_{\rm cur}\;\,(1 - \kappa_{\rm cur} \; I^2\,)\,A^{1/3} 
      B_{\rm cur}({\rm def})
                                                                  \nonumber \\
     &+&
      b_{\rm curG}\,(1 - \kappa_{\rm curG} I^2\,)\,A^0                 
                                                                  \nonumber \\
     &+& 
      \frac{3}{5} \, e^2 \frac{Z^2}{r_0^{ch} A^{1/3}}\, B_{\rm Coul}({\rm def}) 
      - 
      C_{4}\frac{Z^2}{A} 
                                                                  \nonumber \\
     &+& 
      E_{\rm micr}(Z,N;{\rm def}) 
      + 
      E_{\rm cong}(Z,N)\,,
                                                                  \label{ELD}
\end{eqnarray}
where
\begin{equation}
      E_{\rm micr} 
      = 
      E_{\rm pair} + E_{\rm shell} 
                                                                  \label{mic}
\end{equation}
is the microscopic energy containing the contributions from paring and shell 
effects coming from the protons and from the neutrons. The congruence energy 
according to Ref. \cite{MS96} is equal to:
\begin{equation}
      E_{\rm cong} = - 10 \,{\rm MeV} \cdot  \exp ( -42 \, |I|/10) \;.
                                                                  \label{con}
\end{equation}
The term proportional to $Z^{2.39}$ describes the binding 
energy of electrons. The surface diffuseness of the charge distribution
reduces the Coulomb energy proportionally to $Z^2/A$. 

\begin{figure}
  \begin{center}
    \leavevmode
    \epsfig{file=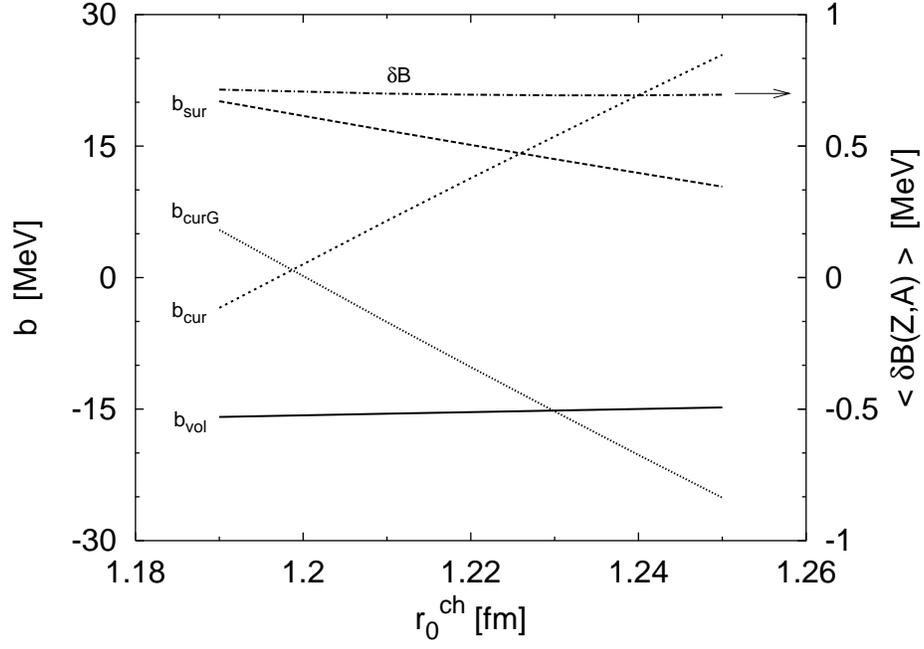, width=12.5cm, angle=00}
  \end{center}
  \caption{Dependence of various liquid drop model terms obtained by the least
           square fits to the experimental masses as functions of the Coulomb 
           radius constant \protect{$r_0^{ch}$}. The corresponding r.m.s. 
           deviation of the differences between the theoretical and experimental
           binding energies $<\delta B>$ refers to the right-hand side 
           ordinate axis.}
  \label{fig03.lsd}
\end{figure}

In order to investigate the interplay between the Coulomb and nuclear energies 
when trying to reproduce the nuclear binding energies we have performed a 
test fit to the experimental data from Ref. \cite{An01} for various choices of 
the charge radius constant $r_0^{ch}$. The results are presented in
Fig.~\ref{fig03.lsd}, where several terms of the liquid drop model are plotted 
as functions of $r_0^{ch}$. The root-mean-square deviation of the binding 
energies, $<\delta B>$, is shown referring to the right-hand side vertical axis. 
Surprisingly, the quality of the fit depends only slightly on the choice of 
$r_0^{ch}$ but the magnitudes of the first and of the second order curvature 
terms change dramatically with $r_0^{ch}$. It is seen that for 
$r_0^{ch}\approx 1.2$ 
fm both curvature terms are small since they both change sign near the above   
$r_0^{ch}$-value.  (This Figure is  similar to the previous 
one (Fig.~\ref{fig02.lsd}) where the dependence of the surface and
curvature terms on the radius of the leptodermous expansion was studied.)

The results in Fig.~\ref{fig03.lsd} show that it is rather 
difficult to fix the Coulomb radius parameter from the binding energies since
the corresponding dependence is a flat function. Trying to deduce the related
curvature contributions when varying both curvature terms is not very easy
either, since the empirical $r_0^{ch}$ value is expected not to differ very 
much from the mentioned special value of about 1.2 fm for which $a_{\rm cur}$ 
and $a_{\rm curG}$ are small (pass both through zero). This is precisely the
type of parametric peculiarity of the macroscopic energy formula that has been
mentioned in Sec.~\ref{Sec01}. Under these conditions the fit to the fission
barrier heights could give a valuable additional criterion. In the next
Sections we are going to present the results of the fit of the parameters of
the traditional (i.e without the curvature terms) liquid drop model energy
expression to the experimental masses and the liquid drop model with the 
curvature
terms where the parameters are adjusted either to both the measured ground 
state masses and fission barrier heights, or to the measured ground state 
masses only.


\subsection{New Parameters of the Traditional Myers-\'Swi\c{a}tecki
            Liquid-Drop Energy-Expression}
\label{Sect03b}            

Exactly the same mass expression as the one of Myers-\'Swi\c{a}tecki liquid 
drop (MS-LD) of Ref.~\cite{MS67}:
\begin{eqnarray}
       M(Z,N;{\rm def} ) 
      &=&  
         Z M_{\rm H} 
       + 
         N M_{\rm n} 
       - 
         0.00001433 \, Z^{2.39}                                   \nonumber \\ 
      &+& 
         b_{\rm vol} \,  (1 - \kappa_{\rm vol} I^2) \, A          \nonumber \\
      &+& 
         b_{\rm surf} \, (1 - \kappa_{\rm surf} I^2) \, A^{2/3}  
                                                                  \nonumber \\
      &+& 
         \frac{3}{5} \, \frac{e^{\,2} Z^2}{r_0^{ch} \, A^{1/3}} 
       - 
         C_{4} \frac{Z^2}{A} 
                                                                  \nonumber \\
      &+& 
         E_{\rm def}(Z,N) 
       + 
         E_{\rm pair}(Z,N) 
       + 
         E_{\rm shell}(Z,N) 
       + 
         E_{\rm cong}(Z,N)\;,
                                                                  \label{MSELD}
\end{eqnarray}
but with the microscopic corrections for deformation, pairing and shell effects
treated as in Ref.~\cite{MN95} and the new estimate of the congruence energy
($E_{\rm cong}$, Ref.~\cite{MS96}) was used to obtain the best fit to the 2766 
empirical binding energies from Ref.~\cite{An01} of the isotopes with the 
proton and neutrons numbers larger or equal to 8. Following a practical recipe 
used in Ref. \cite{MS96}, when adjusting the parameters of the macroscopic
model we take into account the nuclear deformations. In particular, 
the {\em macroscopic part} of the total energy, $E_{\rm def}$, is taken 
from Tables of Ref.~\cite{MN95} ($E_{\rm def}$ is defined as the difference 
between the macroscopic energy of a nucleus at the equilibrium deformation 
and the energy of the same but spherical nucleus, plus the sum of the shell
and pairing energies taken at the actual equilibrium deformation).
The same approximation is used when fitting the parameter sets of other variants 
of the model presented in this paper.

The new set of parameters obtained by fitting the nuclear masses (but not using
any information about the fission barriers, similarly as in Ref.~\cite{MS67}), 
is given below. For comparison the old values of the parameters taken from the
above reference are given in parentheses: 
\bnll{compar}
     && 
        b_{\rm vol} ~= -15.8484    \quad (-15.667) ~ {\rm MeV}\,,                                                                 
                                                           \label{compar.a} \\
     && 
        b_{\rm surf} = 19.3859     \quad  (18.56)  ~ {\rm MeV}\,, 
                                                           \label{compar.b} \\
     && 
        \kappa_{\rm vol} ~= 1.8475 \quad (1.79) \,,      
                                                           \label{compar.c} \\
     && \kappa_{\rm surf} = 1.9830 \quad (1.79) \,,
                                                           \label{compar.d} \\
     && 
        r_{0}^{ch} ~~= 1.18995     \quad  (1.2049)  ~ {\rm fm}\,, 
                                                           \label{compar.e} \\
     && C_{4} ~~= 1.19949          \quad  (1.21129) ~ {\rm MeV}.  
                                                           \label{compar.f}
\enll
The r.m.s. mass deviation corresponding to the new set of parameters and the 
microscopic corrections from Ref.~\cite{MN95} is $<\delta M> = 0.732$ MeV; 
an analogous quantity for the old set of the liquid drop parameters and the 
same microscopic corrections is $<\delta M> = 4.477$ MeV. 
The r.m.s. mass deviation obtained with the new parameter set is comparable 
with the one of the 
Thomas-Fermi model ($<\delta M> = 0.757$ MeV) and proves that the 
liquid drop approximation can reproduce the nuclear masses with a 
comparably high accuracy. 

Let us observe that {\em neither} the old set of the liquid drop parameters 
(MS-LD) {\em nor} the new one (LDM) is able to reproduce correctly the 
magnitudes of the experimental fission barriers. The discrepancies between 
theoretical and experimental fission barrier heights of 40 nuclei
that can be found in the published literature\footnote{In this paper we use
only those experimental barrier heights that can be found in the published
sources; they correspond to 40 nuclei with $75 \leq A \leq 252$. 
This information concerns four relatively light nuclei {\em viz.}
$^{75}_{35}$Br and $^{90,94,98}_{\qquad 40}$Mo and the whole rest of nuclei 
clearly separated in terms of $Z$ ($Z > 70$). The barriers 
of these four lightest nuclei present the same type of difficulties for all 
the variants of the model, including the one introduced in this paper (LSD). 
As far as the barriers of $Z > 70$ nuclei are concerned, some variants of 
the model describe them very well, some variants are clearly less satisfactory  
(for details see below).} are presented in Fig.~\ref{fig04.lsd} (for the sources
cf. Refs.~\cite{MS96,MS99,JM99} and references quoted there). To extract the 
barrier heights from the experimental data we have used a
similar prescription as the one in Ref. \cite{MS99}, namely we define the 
barrier height as a difference between the liquid drop saddle-point
energy and the ground state energy deduced from the ground-state masses. 
It is seen in Fig.~\ref{fig04.lsd} (top) that the traditional 
Myers-\'Swi\c{a}tecki liquid drop (MS-LD) model {\em overestimates} the barrier 
heights of the lighter nuclei by about 10 MeV and by about 3-4 MeV those of the 
heavier ones. Our new fit of parameters of this traditional
liquid drop model (LDM) overestimates the barrier heights even more
significantly (Fig.~\ref{fig04.lsd}, bottom). Does it mean that the liquid
drop model is unable to reproduce with a more respectable accuracy the
positions of the fission saddle-point energies? In order to answer this 
question we have performed additional tests in which we have made either a
simultaneous fit of the liquid drop model parameters to the experimental masses
and fission barrier heights, or the fits limited to the nuclear masses. 
The results are presented in the next Sections.


\subsection{Liquid Drop Model with Curvature Terms}
\label{Sect03c}

The purpose of the following discussion is to examine the influence of the
two curvature terms introduced earlier through relations (\ref{E9}) and 
(\ref{ELD}). 
We would like to adjust the parameters of the curvature-extended liquid drop 
model both to the huge body of the experimental nuclear binding energies 
{\em known today} and, if necessary, to the experimental fission barrier 
heights - in this context we are going to profit from the observations presented
in Sect.~\ref{Sect03a}. The nuclear mass expression of Eq.~(\ref{ELD}), compared 
to the one by Myers and \'Swi\c{a}tecki in Eq.~(\ref{MSELD}), contains 
the curvature terms of the first and of the second orders.
The fit to the experimental masses and fission-barrier heights will be performed
in three different ways: the one where only the second order curvature term was 
included, another one with the first and the second order curvature terms, and 
finally the one with the first order curvature term only.
In particular it will be shown that taking into account the Gauss-curvature 
(second-order) term which is $A$ and deformation independent but may possibly
introduce a strong dependence on the isospin factor $I=(N-Z)/(N+Z)$, improves 
the quality of the mass fit provided the surface tension and related
coefficients were fitted to the fission barriers. It influences indirectly the 
fission barrier heights through an extra $(Z,N)$-dependence in all other 
simultaneously fitted parameters. 

We proceed to discuss the results of the three variants of the fitting 
procedure separately.


\subsubsection{Gauss-Curvature Term}
\label{Sect03c.1}

In order to study the effect of the Gauss-curvature term alone on the liquid 
drop energy expression we set the first order curvature term to zero,
$b_{\rm cur}=0$, thus assuming for the moment that the barrier heights can be 
described by the competition between the surface and Coulomb contributions only, 
very much like in the traditional liquid drop model approaches. When discussing 
the particular
case of spherical nuclei (but the conclusions drawn apply to some extent to the 
moderately deformed nuclei as well) it was shown, cf. Eqs.~(\ref{E13} - 
\ref{E15}), that if one sets $b_{\rm cur}=0$ then necessarily $b_{\rm curG}\neq 0$.

To fit the parameters of the model in this case, we use the fact that only some
of them influence the fission barriers and we proceed as follows. First,
for each value of the charge radius ($r_0^{ch}$), we fix the surface 
coefficients $b_{\rm surf}$ and $\kappa_{\rm surf}$, by making the least square 
fit to all {\em experimental fission barrier heights} listed in 
Ref.~\cite{MS96}.
Then the charge radius and all other than the surface-tension LDM parameters 
in Eq.~(\ref{ELD}), including the Gauss-curvature term, are adjusted 
by the least square fit to the {\em experimental binding energies} of 2766 
isotopes with $Z,N \ge 8$ taken from Ref.~\cite{An01}.

The parameters of such a {\em 'new' liquid drop} (NLD) formula are listed in 
Table~\ref{tab01.lsd}. The mean square deviation of the theoretical and 
experimental binding energies, $<\delta B> = $0.814 MeV, is only slightly 
larger than that of $<\delta B> = $0.732 MeV, obtained with the re-fitted
parameters of the traditional Myers-\'Swi\c{a}tecki liquid drop model (LDM) as
described in Sect.~\ref{Sect03b}. However, the fission barrier heights are now 
much better reproduced. The r.m.s. deviation of the barrier heights for 
all treated nuclei is $<\delta V_B> = $1.90 MeV while for the LDM we found 
$<\delta V_B> = $7.08 MeV (see in Fig.~\ref{fig04.lsd}). Including
the isospin-dependent Gauss-curvature term improves the agreement with the
experimental barrier heights, nevertheless the corresponding new set of
parameters does not reproduce perfectly the barriers: it is seen in
Fig.~\ref{fig05.lsd}, top, that the barriers of the light isotopes ($A \lt
100$) are {\em overestimated} by about 4 MeV and the barriers of nuclei 
with $A \sim 180$ are {\em underestimated} by about 3 MeV while the barrier 
heights of the heaviest nuclei are {\em overestimated} by 1.5 MeV. Thus our
procedure provides, on the average, an improved fit to the experimental fission
barrier heights but it does not reproduce very well neither $Z^2/A$ nor $A$
dependence of them.

Below we show that a possible remedy is to include the first 
order curvature term.
 

\subsubsection{Both Curvature Terms}
\label{Sect03c.2}

It is known that the light nuclei have saddle points at very elongated
shapes whereas the saddle points in the actinide and trans-actinide nuclei 
correspond to rather compact shapes. The surface and curvature terms depend
on deformation in a very similar way for small and even moderate deformations 
\cite{HM88}, while at large deformations the differences become pronounced. 
This feature will be used to improve the description.

Performing the least square fit to the experimental fission barrier heights
for a fixed charge radius ($r_0^{ch}$) we have obtained the surface, $b_{\rm
surf}$ and $\kappa_{\rm surf}$, and the curvature, $b_{\rm cur}$ and
$\kappa_{\rm cur}$ coefficients, all other parameters being insensitive to the
barriers. The charge radius constant as well as the
rest of the parameters of the deformation independent terms in Eq.~(\ref{ELD})
were obtained as before by the least square fit to the known  experimental
masses of Ref.~\cite{An01}. The r.m.s. deviation from the experimental data
obtained with such a procedure is 0.844 MeV for 2766 masses and only 1.06 MeV 
for the fission barriers. The parameters obtained through this procedure
give a very strong dependence of both curvature terms on the reduced isospin,
i.e. the corresponding $\kappa$-coefficients are large. We find:
$b_{\rm cur} = -8.219$ MeV, $\kappa_{\rm cur} = 38.92$ and
$b_{\rm curG} = 21.82$ MeV, $\kappa_{\rm curG} = 25.0$ . 
This dependence leads to the negative first order curvature contribution
for the light nuclei ($A\lt 130$) (recall that the corresponding contribution
is $b_{\rm cur}\,(1 - \kappa_{\rm cur} I^2)$, and thus for $I^2$ small, the
total contribution of this term is negative). 

The next attempt, while using the fitting procedure that employs both 
curvature terms was to fit all 10 parameters of the model, Eq.~(\ref{ELD}),
to the experimental binding energies {\em only}. This lead to the r.m.s. 
deviation from the experimental masses equal to 0.693 MeV, but the fission
barriers obtained in this way were up to 20 MeV too high for the light nuclei
with $A<100$ while for the heaviest nuclei they were by about 2 MeV too
small. These unsatisfactory results lead us to examine more thoroughly the 
use of the first order curvature term only i.e. by setting by definition the 
Gauss curvature term to zero, as discussed in the following section.


\subsubsection{First Order Curvature Term and the LSD Parameter Set}
\label{Sect03c.3}

It turns out that the liquid drop model which in addition to the volume,
surface and Coulomb terms contains only the first order curvature term 
gives the most satisfactory results, as presented below. The parameters of this
{\it Lublin-Strasbourg Drop} (LSD) variant of the macroscopic model are fitted 
solely to the nuclear masses and not to the fission barriers. 
The LSD parameters obtained by fitting to the 2766 experimental masses of 
Ref.~\cite{An01} are listed in Table~\ref{tab01.lsd}. The differences between 
the theoretical and experimental barrier heights are presented in
Fig.~\ref{fig05.lsd}, bottom. Now the mean square deviation of the binding
energies amounts to $<\delta B> = 0.698$ MeV, while the mean square deviation of
the barrier heights $<\delta V_B>$ = 3.56 MeV, but it decreases to only 0.88
MeV when the four lightest nuclei are disregarded i.e. when only the nuclei with 
$Z > 70$ are considered.

As it is seen the parameterization of the barrier heights for heavier nuclei
with $Z > 70$ is improved considerably. The fission barriers obtained with
the LSD model are closer to the experimental ones as compared to analogous
results obtained in Ref.~\cite{MS96} with the Thomas-Fermi model (MS-TF); 
this is illustrated in Fig.~\ref{fig06.lsd}, top. The difference between the 
MS-TF and the measured barriers are plotted in the bottom part of 
Fig.~\ref{fig06.lsd}. It is seen that for heavier nuclei the agreement between 
the experimental data and the LSD fission barriers (Fig.~\ref{fig05.lsd}, 
bottom) is even better than that for
the MS-TF model while for the light isotopes ($A < 100$) both models give
comparable fission barriers, approximately 10 MeV too high.
This large discrepancy between the theoretically predicted fission barrier
heights and the measured values for light nuclei could originate from the
fact that these fission barriers are very broad and the saddle points are
very close to the scission points. At such configurations it could happen that
the negative congruence energy (nearly) doubles, as suggested in 
Ref.~\cite{MS96}, and as a consequence the fission barrier heights calculated 
within such an approach could get much closer the experimental ones; here we do 
not examine this type of effects since the microscopic origin origin of the
congruence effects exceeds the framework of the classical model. 

The role of the curvature term together with its dependence on isospin needs to
be still analyzed in more detail. We shall examine the above questions in the 
next Section.

The calculated LSD masses of 2766 nuclei are compared with the measured ones 
in Fig.~\ref{fig07.lsd}. The lines join the points corresponding to the 
common-isotope chains. 
A part of the observed local discrepancies may originate from 
the microscopic corrections to the macroscopic energies that were evaluated in 
Ref.~\cite{MN95} assuming the same deformations for the proton and neutron 
distributions. The self-consistent calculations made in Ref.~\cite{PR97,WN98}
show that in the ground state the proton and the neutron distributions are not
equally deformed. A rough estimate made in Ref.~\cite{BP00} within the
Hartree-Fock-Bobolubov approximation with the Gogny force shows that this
effect can change the ground state energy by approximately $\pm 0.5$ MeV. The
effect of deformations that are different for the proton and neutron 
distributions can be incorporated to the macroscopic-microscopic models by 
introducing an additional term; this aspect is not going to be developped
in the present paper. 
The form and magnitude of the term responsible for the change of the
macroscopic energy due to the deformation difference of both kinds particles
was estimated in Ref.~\cite{DP02} within the extended Thomas-Fermi model with
the Skyrme forces.

To estimate the 'performance stability' of a given parameter-fit it is 
instructive to examine, among others, how a given mass formula
fitted to a certain 'narrow' mass range performs in an extended mass range
and {\em vice versa}.
For instance, with the LSD parameter set fitted to 1654 isotopes from the
Audi-Wapstra tables we may predict the 2776 masses corresponding to the
compilation of Anthony and by taking the corresponding differences we may
calculate the implied r.m.s. deviations that illustrates the 'predictive power'
of the model and its parametrisation. Such a comparison is presented
in Table~\ref{tab02.lsd} for the LSD parameter set as well as for two other 
models indicated. For comparison also an inverse test has been examined 
i.e. estimating the performance quality when going from a broader mass range
to a narrower one. Results in Table~\ref{tab02.lsd} indicate among others a
remarkable stability or 'predictive power' of the LSD approach: by fitting
the parameters to the 1654 masses and predicting the result for the 2776 masses,
we obtain the r.m.s. deviation of 0.711 MeV, i.e. only 13 keV worse than
the direct fit to the 2776 masses, the latter giving the r.m.s. of 0.698 MeV.


\section{Fission barriers and properties of the potential energy surfaces around
         the saddle points}
\label{Sect05}         
 
Fission barrier heights obtained with the help of the curvature-dependent 
liquid drop (LSD) model have been compared with the experimental data and with 
the estimates of the Myers-\'Swi\c{a}tecki liquid drop (MS-LD) formula 
\cite{MS67}, in Figs.~\ref{fig04.lsd} and \ref{fig05.lsd}. In both models the
fission barrier heights were defined as differences between the LD masses at the
saddle point deformations and the ground state masses. Such a
prescription for the barrier heights was used also in Refs.~\cite{MS96,MS99},
where the authors argued (the so called 'topological property') that the shell
corrections at the saddle points should be small. From Fig.~\ref{fig05.lsd}
it is seen that the fission barriers obtained with the LSD parameter set are
very close to the measured values taken from Refs.~\cite{MS96,JM99,MS99} and
references cited there, but
also that they are systematically overestimated for nuclei with $Z>70$ as it 
was the case of the traditional MS-LD, Ref.~\cite{MS67}, 
model. The LSD barrier heights are also close to those obtained within the
Thomas-Fermi model of Myers and  \'Swi\c{a}tecki what is seen in
Fig.~\ref{fig06.lsd}, top.

Such an agreement was possible only due to the presence of the first order
curvature term in the LSD formula. As it can be seen from Table \ref{tab01.lsd}, 
the LSD curvature energy grows with reduced isospin $I=(N-Z)/A$ while the 
magnitude of the volume and surface terms decreases with $I$. 
Without such an $I$-dependence of the surface and curvature terms 
it would have not been possible to reproduce the whole systematics of the 
barrier heights (it will shown below, Fig.~\ref{fig11.lsd}, that such a
strong dependence is very similar in the ETFSI approach indicating that this
particular result of the fit should not be taken as surprise but rather as an
argument of the physical correctness of the LSD approach). 

One extra remark will be in place here corresponding to the magnitude of the 
curvature coefficient (see also comments related to Eq.~(\ref{Taylb})). Our
parameterization in terms of $b_{\rm cur}$ corresponds more to the discussions
usually associated with the LD energy expression rather than the one used in
the microscopic-type approaches. The correspondence between the two type of
parameterizations has the form
\begin{eqnarray}
       b_{\rm surf} \, [1 - \kappa_{\rm surf} \, I^2]\, A^{2/3}
      &+&
       b_{\rm cur} \, [1 - \kappa_{\rm cur} \, I^2]\, A^{1/3}
                                                                \nonumber \\
       \quad &\to& \quad
       a_{\rm surf} \, A^{2/3}
       + 
       [a_{\rm cur} - 2 a_{\rm surf}^2 / K] A^{1/3}\, ,
                                                                \label{compsur}
\end{eqnarray}
where $K$ denotes the nuclear matter incompressibility coefficient and the
corresponding term represents the semi-infinite nuclear matter contribution.
Even if the isospin dependence in the second line in the above relation would
have been taken into account the above expression makes it clear that  the
direct comparison between the curvature coefficients when using these two
parameterizations cannot be made. However, the factor proportional to 
$a_{\rm surf}^2$ is of the order of 3 MeV and for $a_{\rm cur} \approx 7$ MeV 
the corresponding expression will correspond to the fitted $b_{\rm cur}$ values.
This estimate can be further detailed by using some literature results:
Ref.~\cite{MS90} quotes the estimate $a_{\rm cur} \approx 11$ MeV, while
Ref.~\cite{BG84} $a_{\rm cur}$ ranging from 9.52 to about 13 MeV as obtained
with the six representative Skyrme interactions; one may conclude that our fit 
result and the quoted microscopic model results have comparable orders 
magnitude, our numbers being slightly smaller.

It is interesting to compare the fission barrier profiles obtained with
different parameter sets of the liquid drop model. In Fig.~\ref{fig08.lsd} 
the fission barriers obtained with the traditional Myers-\'Swi\c{a}tecki
(MS-LD), with the new Gauss-curvature dependent (NLD), and that with the
first-order curvature term (LSD)
liquid drop models are plotted for $^{232}$Th (top) and $^{240}$Pu (bottom).
It is seen that in spite of the differences in the barrier heights 
the slopes from the saddle to scission points are similar in all three 
approaches. The barriers are plotted as functions of distance $R_{12}$ (in 
$R_0$ units) between the fission fragments. Each barrier point was minimized 
with respect to all even $\beta_\lambda$ deformations with $\lambda\leq 14$.

The neutron number dependence of the fission barriers of Yb isotopes evaluated
with the MS-LD and LSD parameter sets are presented in Fig.~\ref{fig09.lsd}.
This nuclear range is of particular interest for the hyperdeformation
studies and several, so far unsuccessful experimental tests have been 
already attempted. Each curve is drawn up to the deformation point close to the 
scission point. 
It is seen that the LSD barrier heights are a few MeV smaller than those of MS-LD
model and that they grow less significantly with neutron number. Also the MS-LD
barriers are shorter than the LSD ones. The fission barrier profiles and
their correct description together with the saddle-to-scission path-length are
important when studying the properties of e.g. super- or hyper-deformed
nuclei. In this paper we are not going to go into more details leaving the
corresponding discussion to a forthcoming paper. 
Instead we would like to examine and illustrate on some examples the stiffness 
of the potential energy surface with respect to higher multipolarity 
deformations 
for the elongations which are close to the saddle and/or scission configurations.
This aspect is very important in the studies of e.g. multi-path fission 
mechanisms where the shell energies corresponding to the relatively exotic
(e.g. high-multipolarity) deformations may provide competitive fission
mechanisms. Such a problem arises also at high spins and therefore will
also become important for the new generation of the calculations aiming at
the hyper-deformation effect. In Fig.~\ref{fig10.lsd} the cross sections of the 
potential energy surfaces obtained with the MS-LD and LSD approaches on the
one hand, and with the Yukawa-Folded energy expression with parameters from
\cite{MN81} on the other hand, are plotted for $^{172}$Yb at $\beta = 2$
as functions of $\beta_4$ (top), $\beta_6$ (middle) and $\beta_8$ (bottom).
It is seen that the stiffness properties with respect to these deformations 
are almost the same in the case of the first two compared models.
The YF approach cannot distinguish in any significant manner between, say,
$\beta_4 = 0.5$ and $\beta_4=1.0$ (the corresponding energy difference is
smaller than 1 MeV compared to about 5 MeV in the case of the other two
approaches) and varies only weakly in terms of the higher order multipoles.
This very strong indifference of the YF approach with respect to significant
variations of the nuclear surface at strong elongations was considered for 
some time already as a weakness of the latter approach, cf. Ref.~\cite{LJ96}.

In Fig.~\ref{fig11.lsd} the fission barrier heights of several Fm isotopes 
calculated with the LSD and NLD sets of parameters are compared with the
fission barrier heights obtained in Ref.~\cite{TG00} within the extended
Thomas-Fermi model with the Skyrme interaction (ETFSI). It is seen that the
barrier heights obtained with the NLD and LSD parameters are close to each other
for the light Fm isotopes while for the heaviest ones one may notice a 
significant (3 MeV) difference between the two families of the barrier heights.
This decrease of the barrier heights
with  increasing neutron number $N$ obtained in the LSD model for heavy Fm
isotopes is confirmed by the ETFSI results \cite{TG00}. 

The logarithms of the experimental life-times, $T_{sf}$, are plotted for
comparison, in  Fig.~\ref{fig12.lsd}. It is known from the 
macroscopic-microscopic type of calculations that it was almost impossible to 
reproduce the spontaneous fission life time ($T_{sf}$) systematics for the 
chain of Fm isotopes. For the majority of the theoretical calculations, the 
spontaneous fission life times of heavier Fm isotopes are too long while for 
the light and medium-heavy isotopes they are relatively well reproduced.
An attempt in Ref.~\cite{LS99}, within the macroscopic model that contained no 
curvature terms confirmed the existence of the same deficiency. Such a 
discrepancy in the systematics originates probably from too strong 
$N$-dependence of the macroscopic fission barrier heights; a new parametrisation
can be seen as a step into a right direction.


\section{Summary}
\label{Sect06}

We have shown that it is possible to reproduce simultaneously and with a
reasonable precision the ground-state binding-energies and fission barrier
heights of nuclei within the liquid drop model containing the first and/or the
second order curvature terms. Out of three variants of the model discussed
in detail in this paper the one abbreviated LSD (Lublin-Strasbourg Drop) 
offers the highest precision in the description of masses and fission barriers;
it also has a remarkable stability property with respect to extrapolation
from narrower to the broader range of nuclei.  

The traditional (i.e. without the curvature terms) liquid drop model energy 
expression, abbreviated LDM, with the parameters adjusted to the experimental 
masses only, reproduces remarkably well the experimental masses but gives the 
barrier heights about (3 to 15) MeV bigger than their measured values. 

The liquid drop model with parameters fitted simultaneously to the experimental
binding energies and barrier heights can reproduce rather well both types of 
data when it contains the $A$-independent (but isospin-dependent) second order 
curvature (Gauss) term. This almost traditional expression, i.e. without the 
first order  curvature terms, but with the surface tension adjusted to the
experimental  barrier heights, abbreviated NLD, reproduces {\em on the average}
the right  positions of the saddle points but gives a rather poor {\em
systematic  dependence} of the barrier heights on $Z^2/A$. 

The LSD variant of the liquid drop model contains the term  proportional to
$A^{1/3}$ (first-order curvature-term) and no Gauss-curvature term. It can
reproduce the experimental binding energies and the fission  barrier heights
with an accuracy  comparable to- or better than the Thomas-Fermi model of
Ref.~\cite{MS96},  or the HF+BCS model with Skyrme forces of Ref.~\cite{TG00}.
Perhaps surprisingly, its parameters are adjusted to the experimental binding
energies  only - no information about the fission barriers has been used to fit
the LSD variant parameters. Yet, it gives a correct description of the masses
and the fission barriers,
with the performance comparable to or better than that of other models.  It
gives simultaneously the right {\em systematic} of the barrier heights for
the isotopes with $Z > 70$. The most important information about these
results is contained in Tables ~\ref{tab01.lsd} and \ref{tab01.lsd} of the
paper.

Similarly as in the Thomas-Fermi model of Ref.~\cite{MS96} the LSD fission
barriers of the lighter nuclei ($A < 100$) are overestimated by about 10 MeV.
Here our conclusions coincide with those of \cite{MS99} where the concept of the
congruence mechanism has been discussed to remedy this problem. The isospin
dependence of the surface and curvature terms in the LSD energy expression is
qualitatively confirmed by the systematic of the spontaneous fission life
times of Fermium isotopes and quantitatively by the results of the ETFSI
model. 

In parallel with completing this study, an extension of the present
considerations to the case of the nuclear rotation has been examined and
a number of independent tests of performance of the LSD variant of the  model 
through comparison to the measured barrier heights at high angular momenta has 
been advanced. An agreement with the results on fission barriers for a few
rotating nuclei has been found comparable to the one discussed in this
paper for the static case \cite{He02}.


\acknowledgments

One of the authors (K. P.) wishes to express his thanks for the warm hospitality
extended to him by the Institute for Subatomic Research (IReS) and the Louis 
Pasteur University of Strasbourg; he is particularly indebted to the French 
Ministry of National Education for the invitation as {\em P.A.S.T.} guest 
professor that enabled the present collaboration.

\appendix{}
\section{A short summary of the Thomas-Fermi model of Myers and \'Swi\c{a}tecki}
\label{Appendix}

In Ref. \cite{MS96} the Thomas-Fermi model was applied to a nucleus in
which the nucleons interact via modified Seyler-Blanchard \cite{SB61}
forces:
\begin{equation}
      v_{12} 
      = 
      \frac{2T_0}{\rho_0}  \cdot  Y(r_{12})
      \cdot 
      \left[ -\alpha + \beta 
      \left( {p_{12} \over P_0} \right)^2
     -\gamma\,{P_0 \over p_{12}} + \sigma 
      \left({2\bar\rho \over \rho_0} \right)
      \right]\,. 
\end{equation}
Here $T_0$, $P_0$ and $\rho_0$ are the Fermi energy, the Fermi
momentum and the nuclear matter density, respectively.
The function $Y(r_{12})$ describes the  Yukawa and Coulomb interactions:
\begin{equation}
      Y(r_{12}) 
      = 
      \frac{1}{4\pi a^{\,3}} \; 
      \frac{e^{r_{12}/a}}{r_{12}/a} 
      + 
      V_{\rm Coul}\,.
                                                                   \label{yr12}
\end{equation}
and $\bar\rho$ is the average of the densities of particles "1" and "2":
\begin{equation}
      \bar\rho^{\; 2/3} 
      = 
      (\rho_1^{\; 2/3} + \rho_2^{\; 2/3})/2 \; .
                                                                   \label{aver}
\end{equation}
The mass defect is described by the following expression: 
\begin{equation}
      \Delta M(N,Z) 
      = 
      Z M_{\rm H} + N M_{\rm n} + E_{\rm TF} - 0.00001433 \; Z^{\,2.39}
      +
      E_{\rm shell} + E_{\rm pair} + E_{\rm cong} \;. 
                                                                   \label{exce}
\end{equation}
The corrections for the shell, pairing and deformation effects are 
taken from the tables of Moeller \cite{MN95}.
The congruence energy according to Ref.~\cite{MS96} is equal to:
\begin{equation}
      E_{\rm cong} 
      = 
      - 10 {\,\rm MeV}\; \exp( -42 |I|/10) \,\,.$$
                                                                   \label{cong}
\end{equation}
The Thomas-Fermi mass expression depends finally on the range of Yukawa forces 
$a$ and six adjustable parameters $\xi$, $\zeta$, $\alpha$, $\beta$, $\gamma$, 
$\sigma$ {\em via}
\bnll{pars}
      \alpha_{\ell,u}
     &=&
      \frac{1}{2} (1 \mp \xi )   \alpha \,,
                                                              \label{pars.a} \\
      \beta_{\ell,u}
     &=&
      \frac{1}{2} (1 \mp \zeta ) \beta \,,
                                                              \label{pars.b} \\
      \gamma_{\ell,u}
     &=&
      \frac{1}{2} (1 \mp \zeta ) \gamma \,,
                                                              \label{pars.c} \\
      \sigma_{\ell,u}
     &=&
      \frac{1}{2} (1 \mp \zeta ) \sigma \,,
                                                              \label{pars.d} 
\enll
where $\ell,\,u$ refer to 'like' and 'unlike', and are associated with the 
minus and plus signs, respectively.

The least square fit done for 1654 isotopes approximates the nuclear
masses with the r.m.s. deviation $<\delta M> = 0.655$~MeV.


\newpage

\begin{table}[h]
\caption[TT]{The parameters of the liquid drop model fitted to the
             measured atomic masses only (LDM and LSD) and to experimental
             barriers heights and masses (NLD).
\label{tab01.lsd}}
\begin{center}
\begin{tabular}{ccrrr}
        Term         & Units &  LDM$\;$ &  NLD$\;$ & LSD$\;$  \\
\hline
     $b_{\rm vol}$   &  MeV  & -15.8484 & -15.4721 & -15.4920 \\
  $\kappa_{\rm vol}$ &   -   &   1.8475 &   1.6411 &   1.8601 \\
    $b_{\rm surf}$   &  MeV  &  19.3859 &  17.0603 &  16.9707 \\
 $\kappa_{\rm surf}$ &   -   &   1.9830 &   0.7546 &   2.2938 \\
    $b_{\rm cur}$    &  MeV  &        - &        - &   3.8602 \\
 $\kappa_{\rm cur}$  &   -   &        - &        - &  -2.3764 \\
    $b_{\rm curG}$   &  MeV  &        - &  10.3574 &        - \\
 $\kappa_{\rm curG}$ &   -   &        - &  13.4235 &        - \\
        $r_0$        &  fm   &   1.18995&  1.21610 &   1.21725\\
       $C_{4}$       &  MeV  &   1.1995 &   0.7952 &   0.9181\\
\hline
    $<\delta B>$     &  MeV  &   0.732  &   0.814  &   0.698  \\
\hline
   $<\delta V_B>$    &  MeV  &   7.08   &   1.90   &   3.56  \\
\hline
$<\delta V_B>(Z\gt70)$&  MeV  &   5.58   &   1.56   &   0.88
\end{tabular} 
\end{center}
\end{table}

\begin{table}[h]
\caption[TT]{Root means square deviations (in MeV) of the theoretical
             and the experimental binding energies of isotopes with $Z,N \geq 8$.
             The experimental masses are taken from the Audi-Wapstra
             tables (1654 isotopes) and from Anthony compilation 
             (2766 isotopes). In the first column the numbers of experimental 
             masses are indicated as used when fitting the parameters
             for the LSD variant of the present article as well as for the
             Thomas-Fermi and Hartree-Fock with Skyrme parameter set $MSk7$. 
             The second and third columns contain the performance test and
             the 'extrapolation test' for the fits with the numbers of masses
             given in the head of those columns.
\label{tab02.lsd}}
\begin{center}
\begin{tabular}{lcc}
        Model        & r.m.s (2766) &  r.m.s. (1654)     \\
\hline
        LSD(2766)    &  0.698  &  0.610                  \\
        LSD(1654)    &  0.711  &  0.600                  \\
        TF-MS(1654)  &  0.757  &  0.655                  \\
        MSk7(1888)   &  0.828  &  0.738                  \\
\hline
\end{tabular} 
\end{center}
\end{table}

%
%

\newpage

\newpage

\begin{figure}
  \begin{center}
    \leavevmode
    \epsfig{file=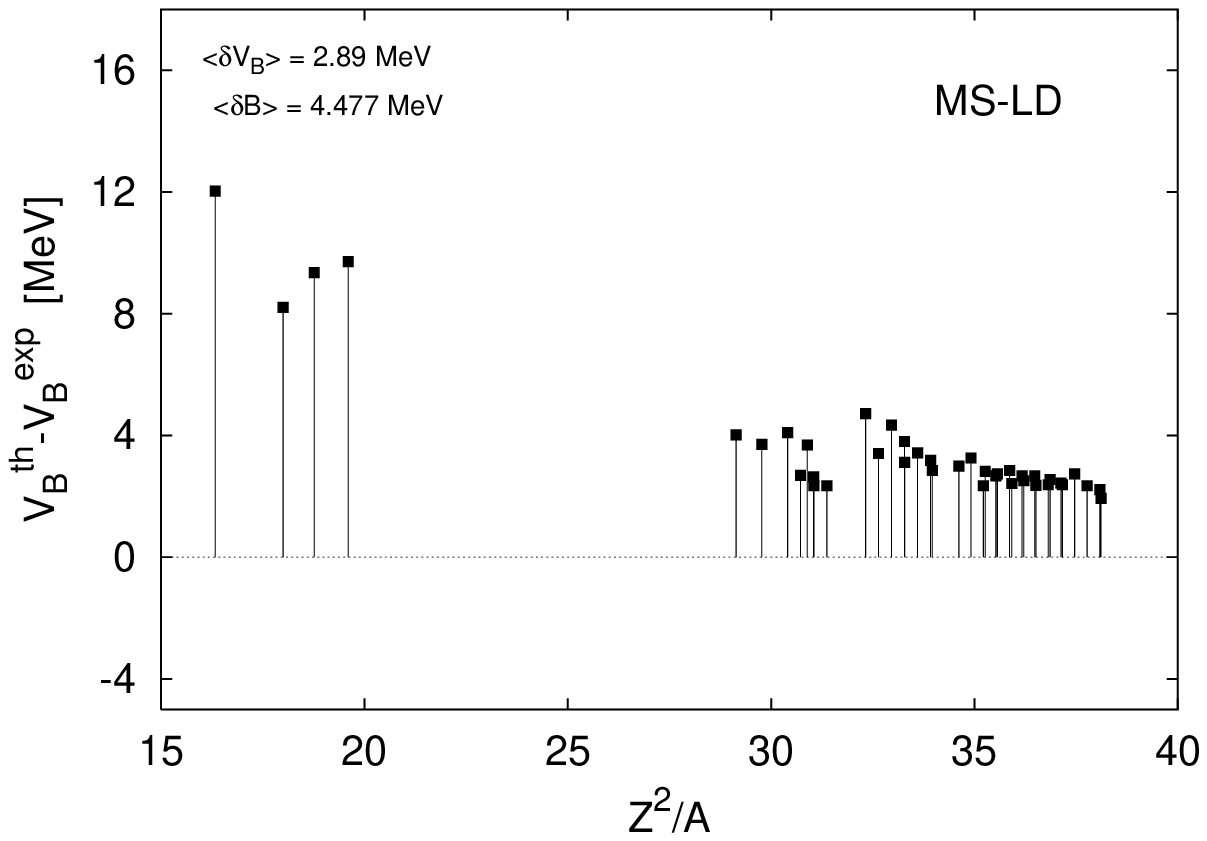, width=12.5cm, angle=00}
    \epsfig{file=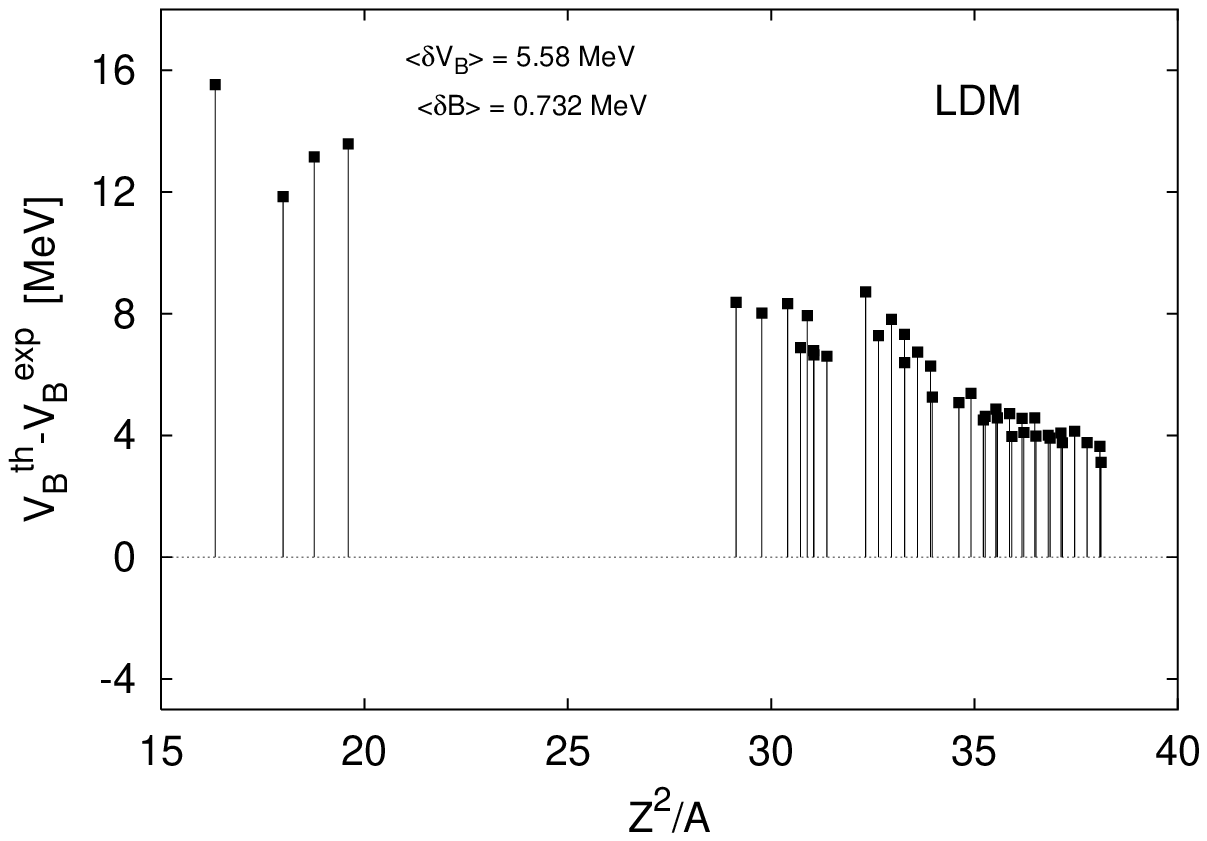, width=12.5cm, angle=00}
  \end{center}
  \caption{The differences between the theoretical and experimental fission
           barriers heights obtained with the traditional Myers-\'Swi\c{a}tecki 
           liquid drop (MS-LD) \protect\cite{MS67}, top, and its modern version 
           (LDM) obtained by the new fit to the {\em presently known} masses and 
           microscopic corrections from Ref.~\protect\cite{MN95}, bottom.
           No information on the barrier heights has been used in the fitting
           procedure in this case.}
  \label{fig04.lsd}
\end{figure}

\begin{figure}
  \begin{center}
    \leavevmode
    \epsfig{file=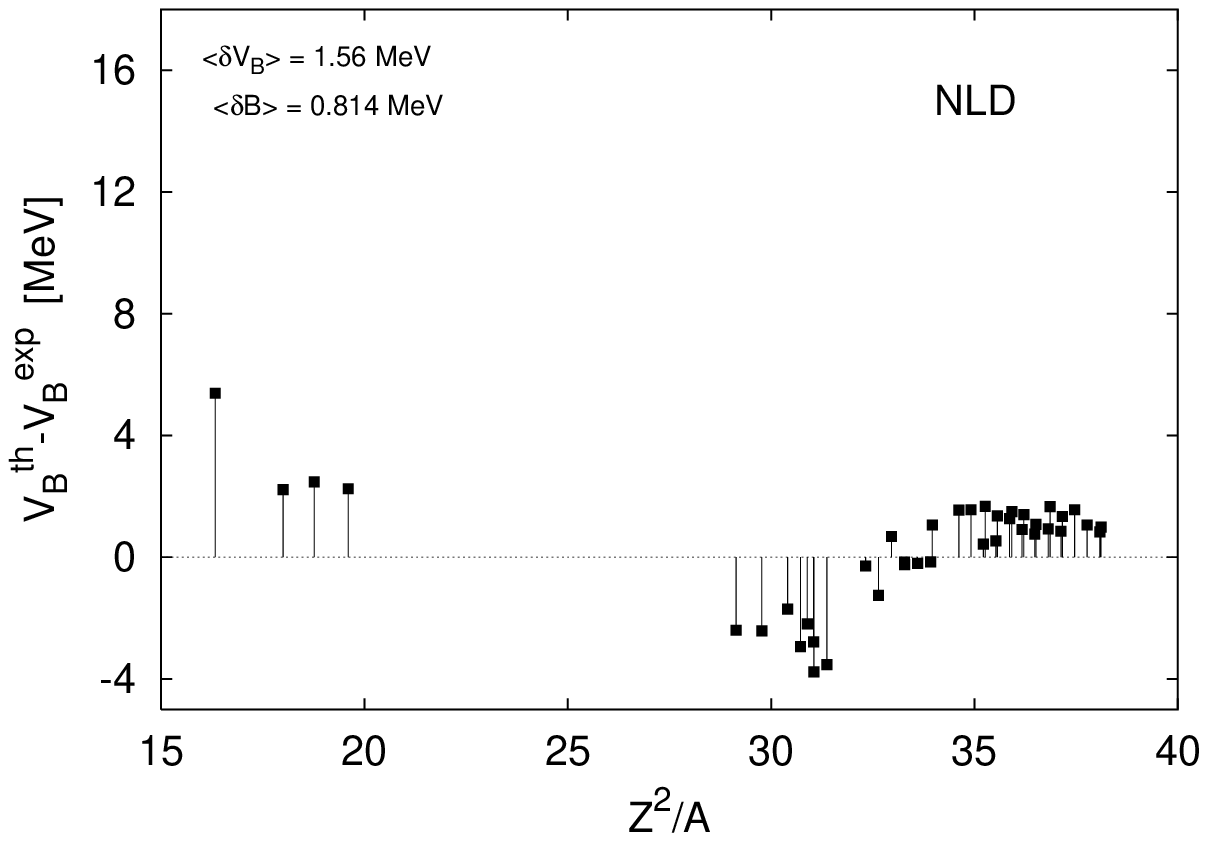, width=12.5cm, angle=00}
    \epsfig{file=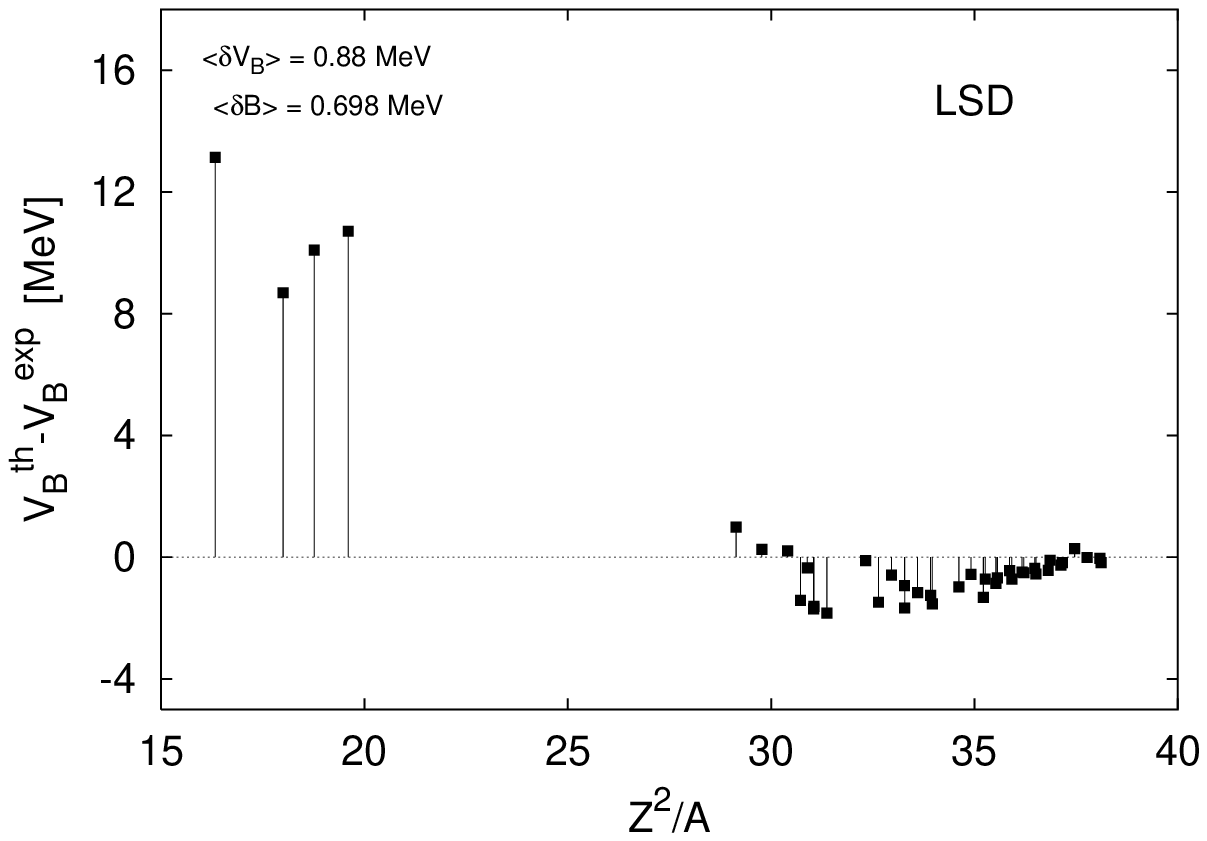, width=12.5cm, angle=00}
    
  \end{center}
  \caption{Differences between the theoretical and experimental fission
           barrier heights obtained with a new liquid drop (NLD) model 
           containing no first order curvature term (top) and with the 
           Lublin-Strasbourg Drop (LSD) model which contains the first order 
           curvature term (bottom). The LSD parameters were adjusted to the
           experimental binding energies only while the NLD ones were fitted to
           the measured fission barrier heights and to the masses.} 
\label{fig05.lsd}
\end{figure}

\begin{figure}
  \begin{center}
    \leavevmode
    \epsfig{file=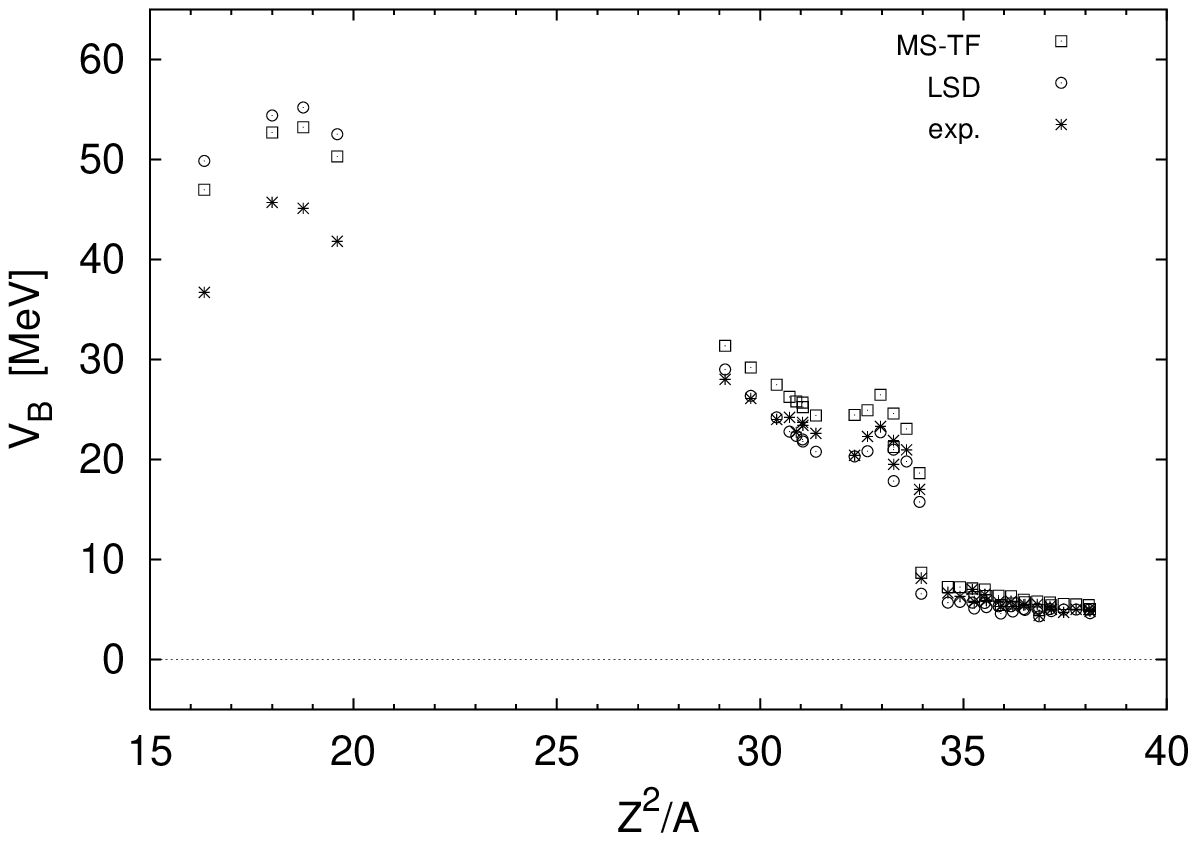, width=12.5cm, angle=00}
    \epsfig{file=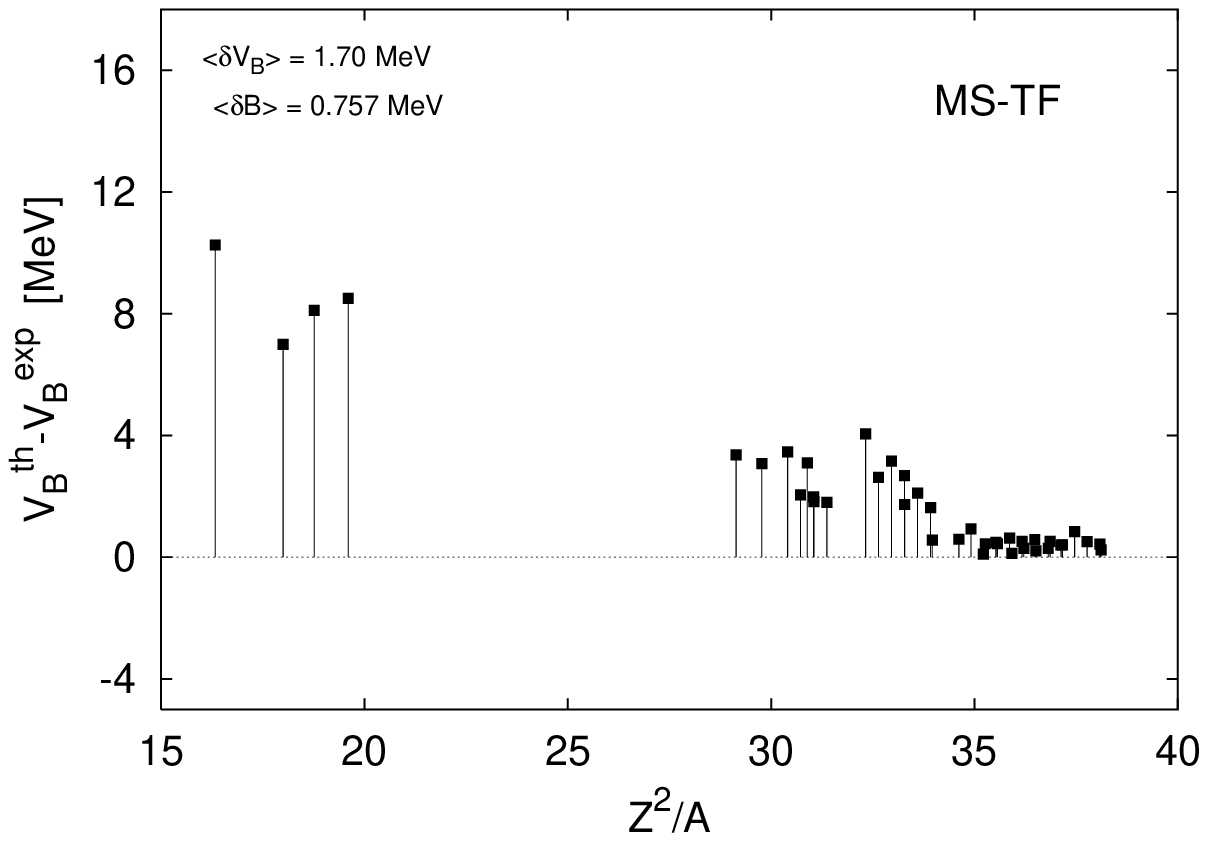, width=12.5cm, angle=00}
  \end{center}
  \caption{Experimental fission barrier heights 
           (see Refs.~\protect\cite{MS96,MS99,JM99} and references cited
           therein), asterisks, compared to the 
           theoretical ones obtained with the LSD (circles) and the 
           Thomas-Fermi models of Ref.~\protect\cite{MS96}, 
           open squares, (top). The differences between the 
           Thomas-Fermi and experimental fission barrier heights are 
           plotted in the bottom diagram.}
  \label{fig06.lsd}
\end{figure}

\begin{figure}
  \begin{center}
    \leavevmode
    \epsfig{file=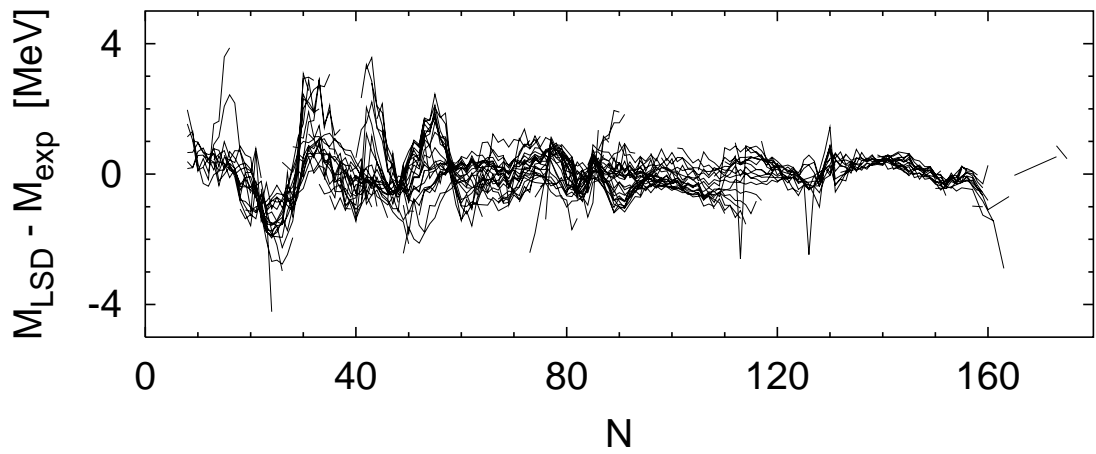, width=15.5cm, angle=00}
  \end{center}
  \caption{Difference between calculated (LSD) and measured (exp.) masses
           for 2766 nuclei from the tables of Anthony. Lines connect the
           isotopes of each given element.}
  \label{fig07.lsd}
\end{figure}

\begin{figure}
  \begin{center}
    \leavevmode
    \epsfig{file=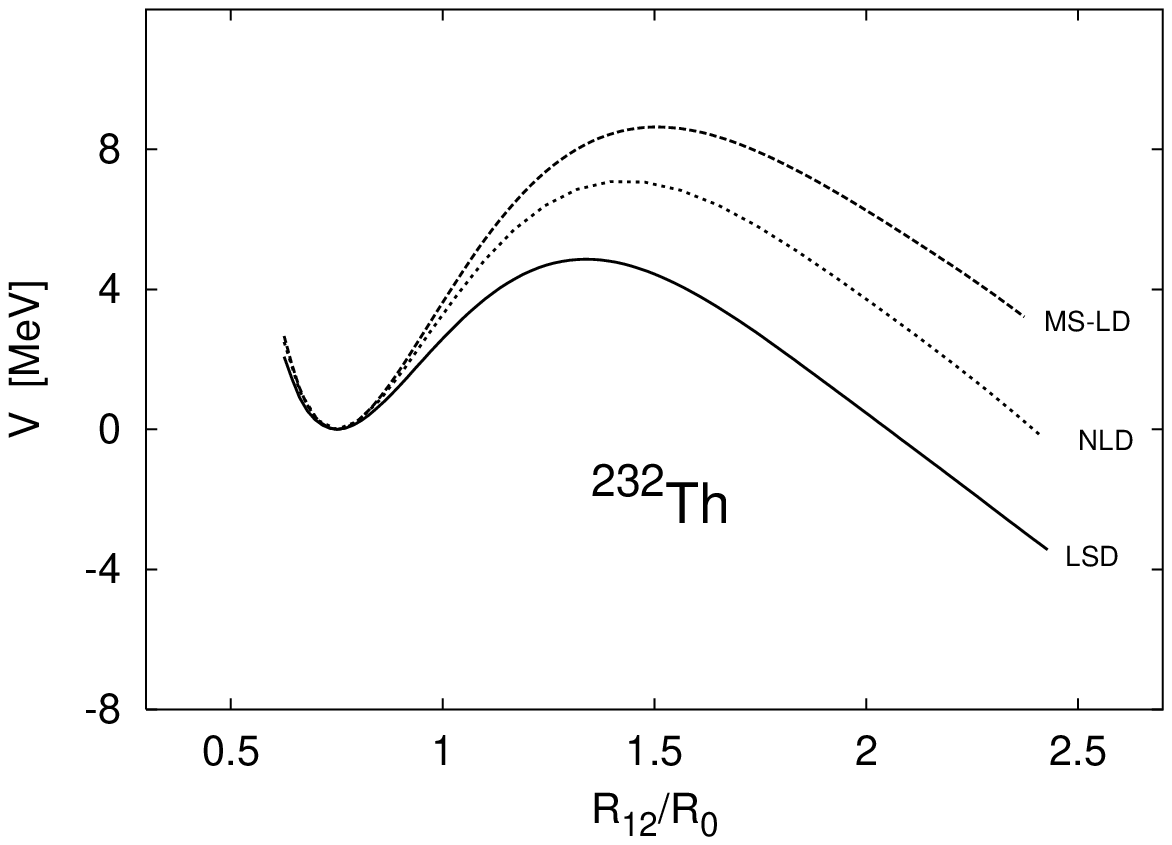, width=12.5cm, angle=00}
    \epsfig{file=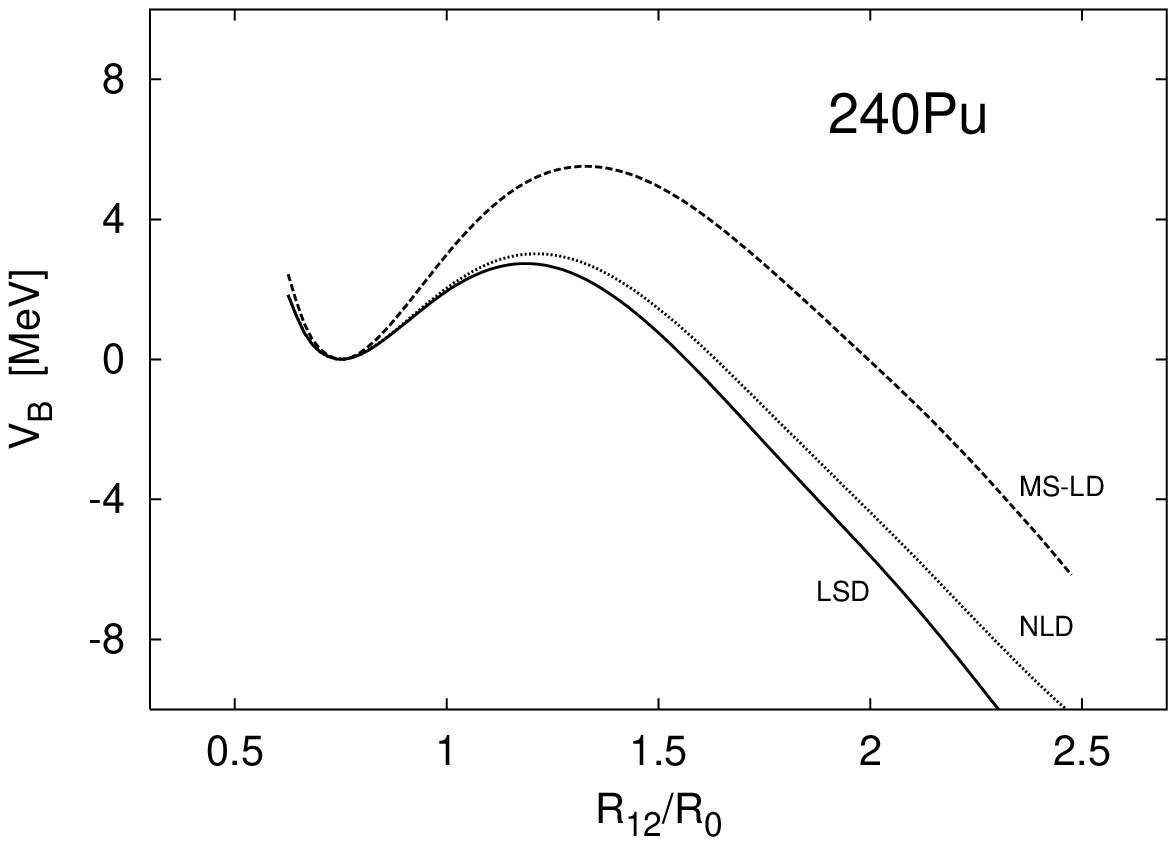, width=12.5cm, angle=00}
  \end{center}
  \caption{Liquid drop fission barriers for $^{232}$Th (top) 
           and $^{240}$Pu (bottom) obtained with the LSD, NLD and MS-LD sets 
           of parameters.}
  \label{fig08.lsd}
\end{figure}

\begin{figure}
  \begin{center}
    \leavevmode
    \epsfig{file=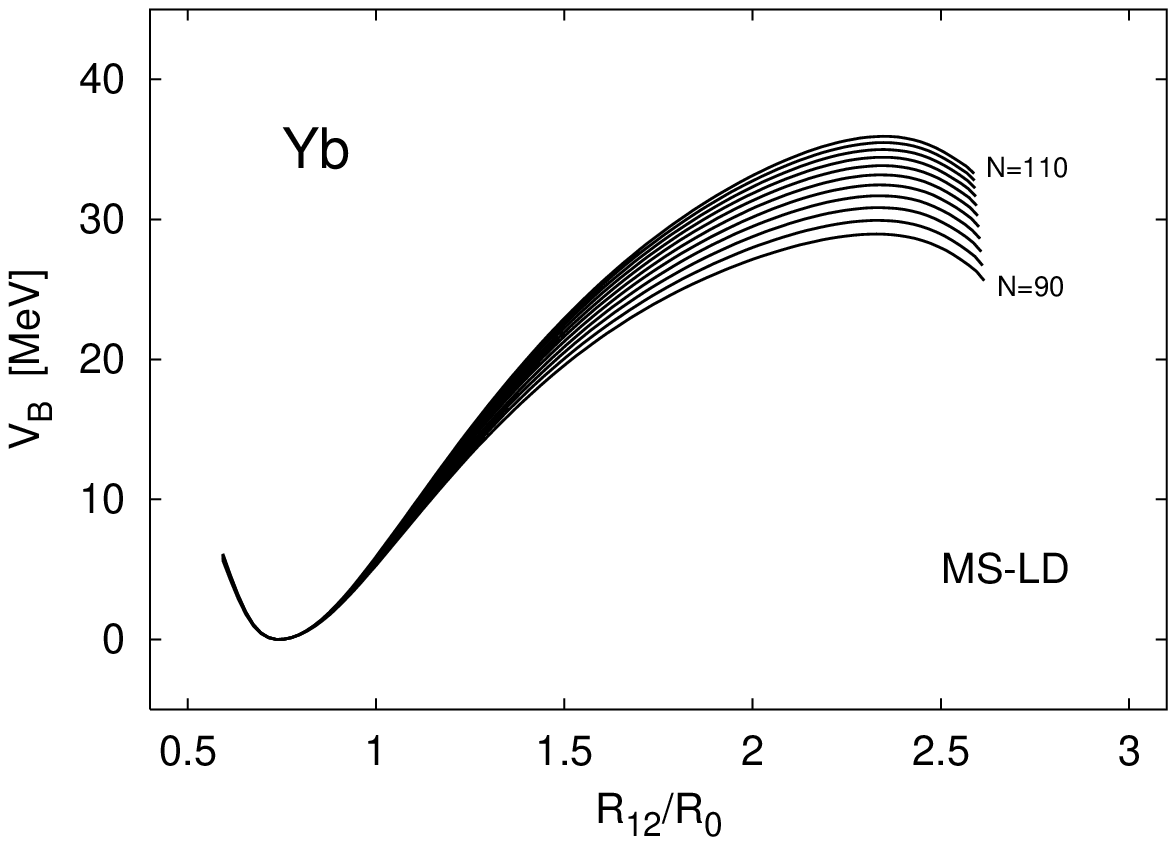, width=12.5cm, angle=00}
    \epsfig{file=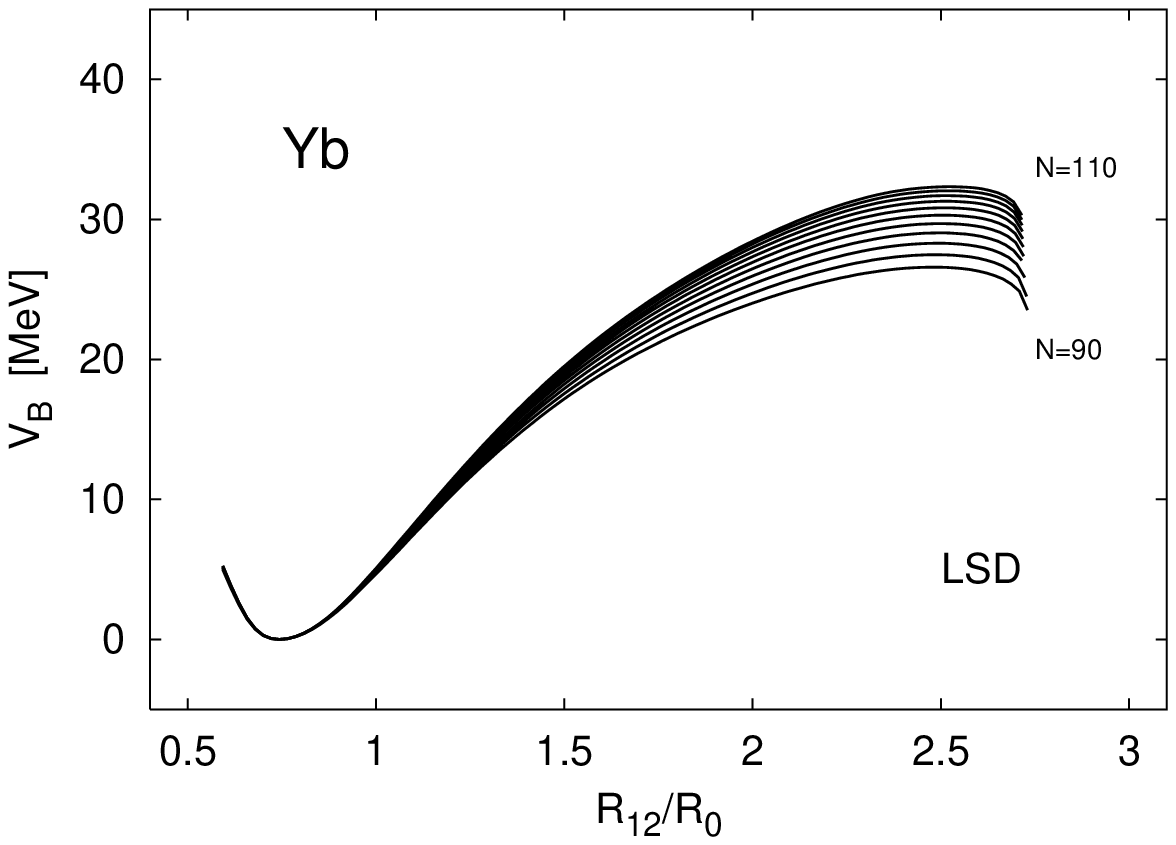, width=12.5cm, angle=00}
  \end{center}
  \caption{Liquid drop fission barriers for Ytterbium nuclei according
           to the traditional (MS-LD) approach (top) 
           and the curvature dependent formulation of the liquid drop
           model with the LSD parameterization (bottom). According to
           earlier predictions the Ytterbium range nuclei are likely
           to be sufficiently stable at the high spins to form the hyperdeformed
           configurations and the corresponding rotational bands.}
  \label{fig09.lsd}
\end{figure}

\begin{figure}
\begin{center}
\leavevmode
\epsfig{file=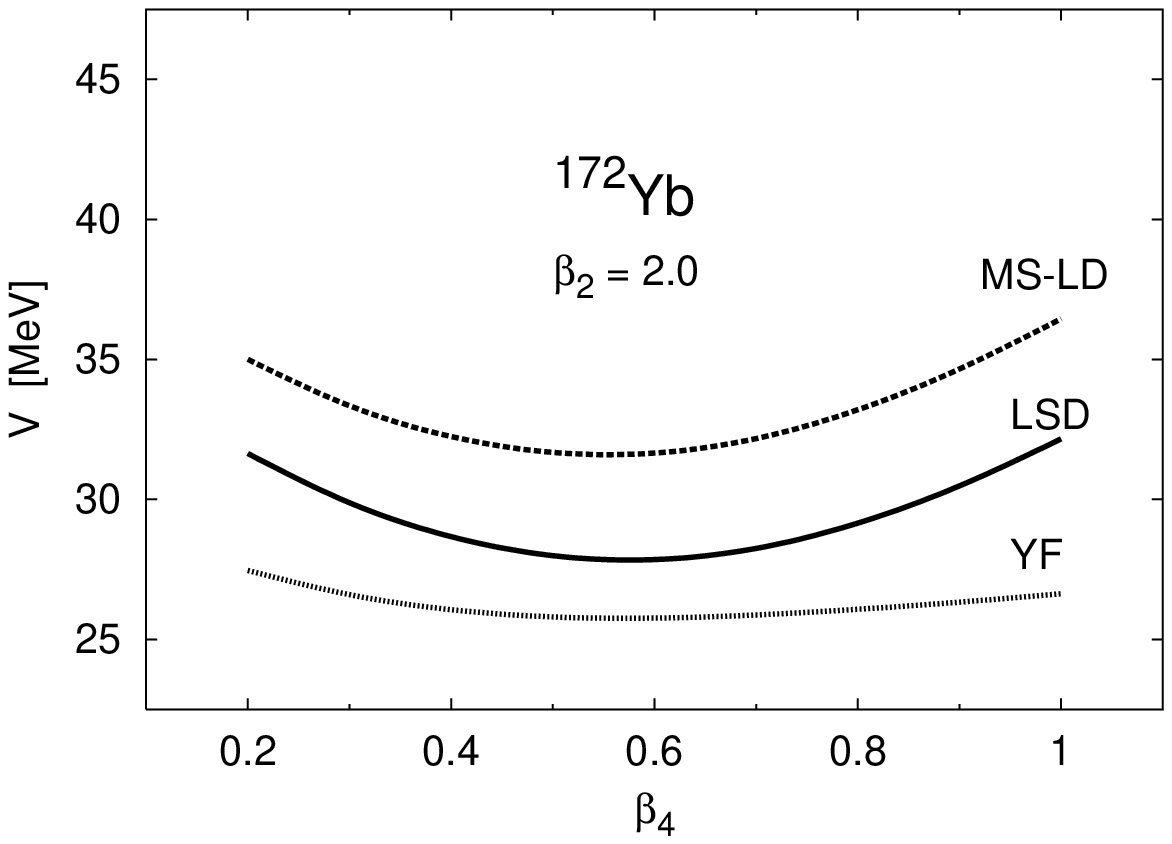, width=12.5cm, height=6.50cm, angle=00}
\epsfig{file=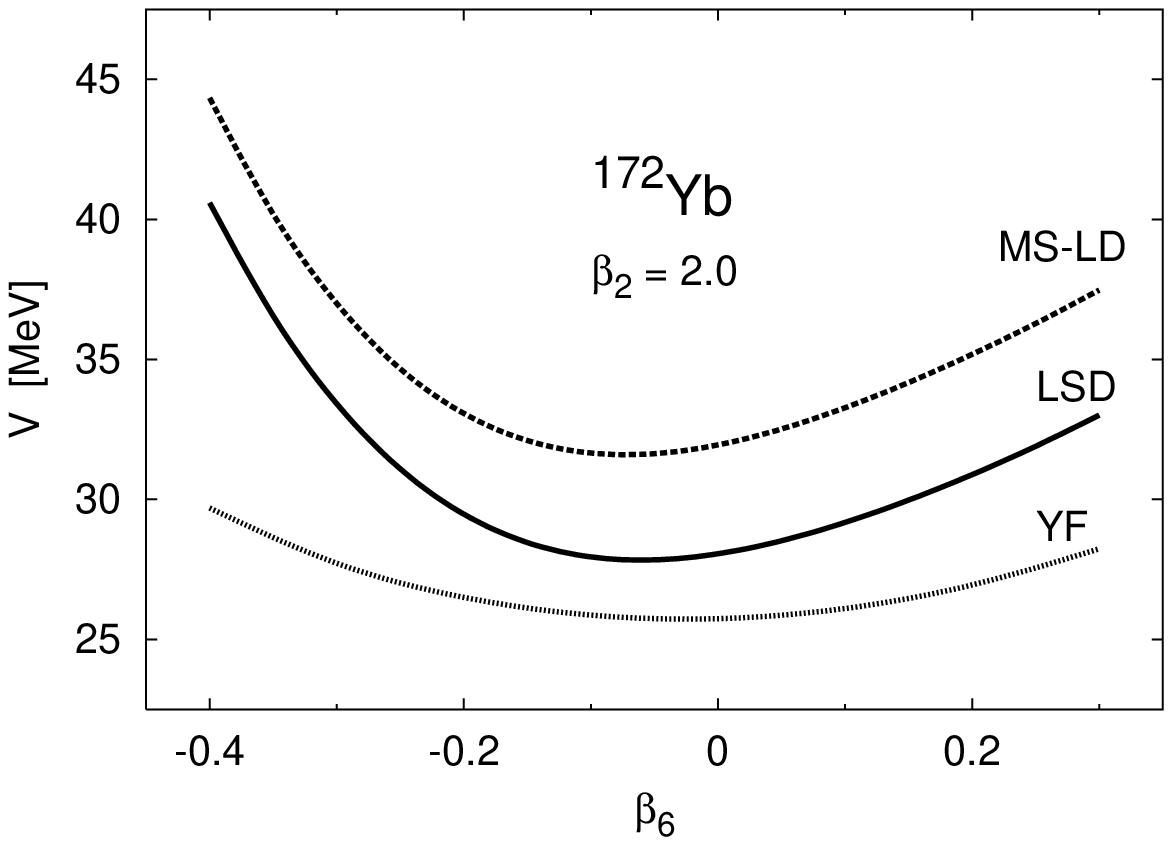, width=12.5cm, height=6.50cm, angle=00}
\epsfig{file=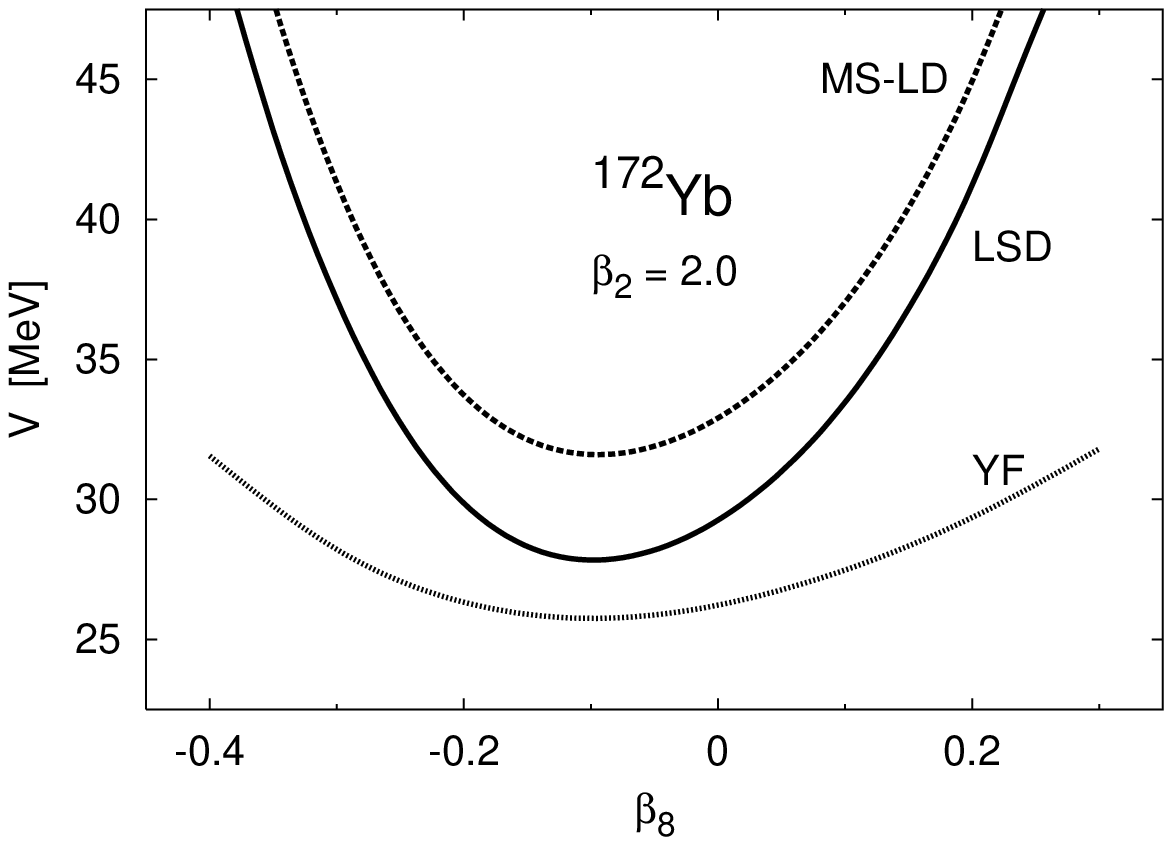, width=12.5cm, height=6.50cm, angle=00}
\end{center}
\caption{Traditional (MS-LD) and curvature dependent (LSD) liquid drop 
         energy of $^{172}$Yb around the saddle point ($\beta_2$=2.0,
         $\beta_4$=0.582, $\beta_6$=-0.058, $\beta_8$=-0.108, 
         $\beta_{10}$=-0.001,
         $\beta_{12}$=0.020) as 
         a function of the deformation $\beta_4$ (top), $\beta_6$ (middle) 
         and $\beta_8$ (bottom). For comparison the Yukawa-Folded (YF) 
         macroscopic model results are shown.}
\label{fig10.lsd}
\end{figure}

\begin{figure}
  \begin{center}
    \leavevmode
    \epsfig{file=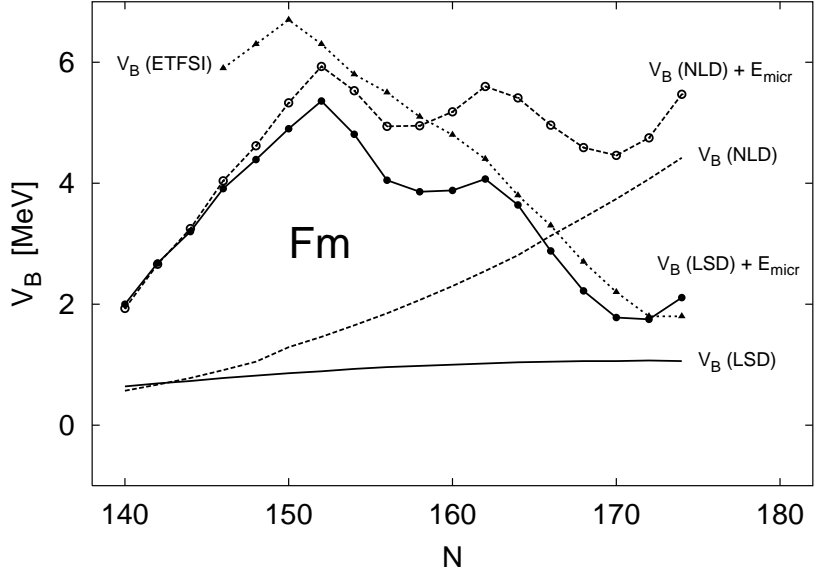, width=11.5cm, angle=00}
  \end{center}
  \caption{Fission barrier heights ($V_B$) of Fermium isotopes evaluated as
          the difference between the liquid-drop saddle-point energy and the
          ground state energy containing the microscopic corrections.
          The solid lines with the full dots correspond to the barriers 
          calculated with the curvature dependent (LSD) model while the dashed
          lines with open circles represent the barriers calculated with the
          liquid drop model without curvature term (NLD). The difference
          between the full and the dotted lines is equal to the ground state
          microscopic correction taken from the tables \protect\cite{MN95}. 
          The fission barriers obtained within the extended Thomas-Fermi model 
          with the Skyrme interaction (ETFSI) \protect\cite{Ma01} are drawn 
          for comparison.}
  \label{fig11.lsd}
\end{figure}

\begin{figure}
  \begin{center}
    \leavevmode
    \epsfig{file=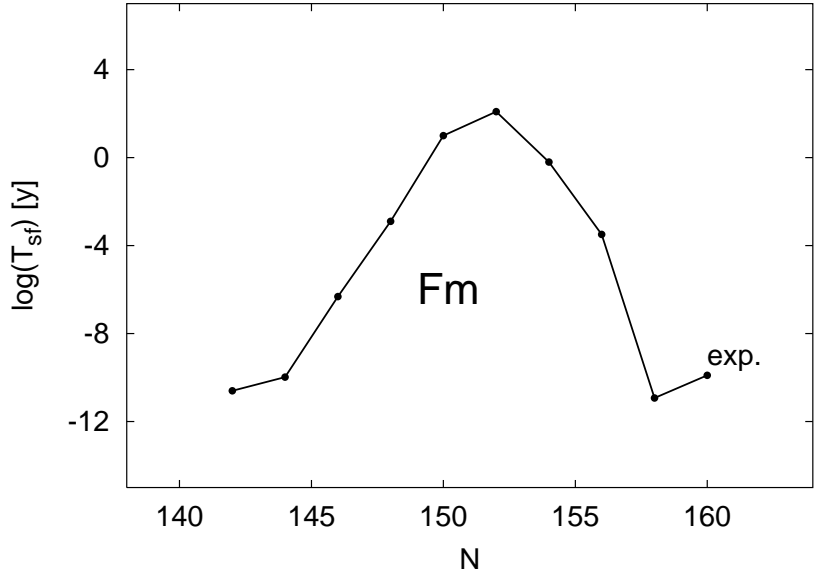, width=11.5cm, angle=00}
  \end{center}
  \caption{Logarithm of the spontaneous fission life time [in years] of Fm
           isotopes. The full dots represent the experimental values. (cf.
           theoretical estimates in Fig.~\protect\ref{fig11.lsd}}
  \label{fig12.lsd}
\end{figure}

\end{document}